\documentclass[8pt,a4paper]{article}
\usepackage{jcappub}
\usepackage{graphicx}

\newcommand{\uvec}[1]{\boldsymbol{\hat{\textbf{#1}}}}
\usepackage{graphicx}
\usepackage{subcaption}
\usepackage{amsmath,amssymb}
\usepackage[font=small,labelfont=bf]{caption}
\usepackage{verbatim}
 \usepackage{url}
 
{} 

\def\araa{\ref@jnl{ARA\&A}}

\title{Detectability of the $\tau_{\rm es}-$21cm cross-correlation:\\ a tomographic probe of patchy reionization}
\author[a,d,h]{Anirban Roy,}
\author[a,b,c,d]{Andrea Lapi,}
\author[e,f]{David Spergel,}
\author[g]{Soumen Basak,}
\author[a,b,c,d]{Carlo Baccigalupi}
\affiliation[a]{SISSA, Via Bonomea 265, 34136 Trieste, Italy}
\affiliation[b]{INFN-Sezione di Trieste, via Valerio 2, 34127 Trieste, Italy}
\affiliation[c]{INAF-Osservatorio Astronomico di Trieste, via Tiepolo 11, 34131 Trieste, Italy}
\affiliation[d]{IFPU: Institute for fundamental physics of the Universe, Via Beirut 2, 34014 Trieste, Italy}
\affiliation[e]{Department of Astrophysical Sciences, Princeton University, Princeton, NJ 08544, USA}
\affiliation[f]{Center for Computational Astrophysics, Flatiron Institute, 162 5th Ave, New York, NY 10003, USA}
\affiliation[g]{School of Physics, Indian Institute of Science Education
and Research Thiruvananthapuram, Maruthamala PO, Vithura,
Thiruvananthapuram 695551, Kerala, India}

\affiliation[h]{Department of Astronomy, Cornell University, Ithaca, NY 14853, USA}

\emailAdd{aroy@sissa.it}

\abstract{The cross-correlation between fluctuations in the electron scattering optical depth $\tau_{\rm es}$ as probed by future Cosmic Microwave Background (CMB) experiments, and fluctuations in the 21cm differential brightness temperature $\Delta T_{\rm 21cm}$ as probed by ground-based radio interferometers, will trace the reionization history of the Universe. In particular, the $\tau_{\rm es}-$21cm cross-correlation should yield a determination of the characteristic bubble size distribution and ionization fraction as a function of redshift. When assuming that the cross-correlation signal is limited by instrumental noise rather than by foregrounds, we estimate its potential detectability by upcoming experiments. Specifically, the combination of HERA and Simons Observatory, CMB-S4 and PICO should yield a signal-to-noise ratio around 3 - 6, while and the exploitation of the SKA should increase it to 10-20. Finally, we have discussed how such levels of detectability can be affected when (simply modeled) 21cm foregrounds are present. For the most promising PICO$\times$SKA configuration, an efficiency of foreground removal to a level of $7\times 10^{-4}$ is needed to achieve a $5\sigma$ detection of the cross-correlation signal; in addition, safe avoidance of foreground contamination in the line-of-sight Fourier modes above $0.03 \,h\,\rm Mpc^{-1}$ would guarantee a detection significance around $3\sigma$.}


\begin{document}
\maketitle
\flushbottom
\section{Introduction}\label{sec|intro}
Reionization leaves its imprints on the cosmic background radiation (CMB) through distinctive signatures in polarization on both small and large scales \cite{Hu2000, Chen2017,Millea2018}, and through a suppression of small scale temperature fluctuations \cite{Planck2016, Planck2018}. The recent measurements from state-of-the-art cosmological experiments have provided increasingly tight constraints on the line-of-sight (l.o.s.) integrated optical depth of electron scattering: $\tau_{\rm es}=0.058\pm 0.012$ and $\tau_{\rm es}=0.054\pm 0.0073$ according to the 2016 and 2018 \textit{Planck} data releases \cite{Planck2016,Planck2018}, respectively; these measurements suggest that the reionization process has been completed at redshift $z\sim 6-8$ but cannot constrain neither its detailed physics nor its accurate temporal evolution and duration.

Recent determinations of the UV luminosity functions out to $z\sim 10$ strongly suggest that faint high-redshift galaxies constitute the main sources of cosmic reionization. \citet{Lapi2017} have shown that a reionization history consistent with \textit{Planck}'s optical depth measurements and with several other independent astrophysical observables (e.g., Lyman-$\alpha$ forest transmission profiles, sizes of quasar near zones, gamma ray bursts (GRB) damping wing absorption profiles, abundance and clustering of Lyman-$\alpha$ emitters, size evolution of Lyman-$\alpha$ halos, photoionization rates inferred from quasar proximity effect) can be naturally obtained
by integrating the observed galaxy luminosity function down to a UV magnitude limit $M_{\rm UV}\sim -13$ and assuming a standard initial mass function as well as reasonable values of the escape fraction $f_{\rm esc}\lesssim 5-10\%$ for ionizing photons from primeval galaxies. Additional sources like Active Galactic Nuclei (AGNs) and quasars may provide a minor contribution, since the number density of the bright ones falls quickly above $z>6$  and the ionization power of fainter ones is insufficient to ionize the InterGalactic Medium (IGM), unless extreme values of the associated escape fraction (around $100\%$) are adopted \cite{Barkana2001,Robertson2010}.

During the epoch of reionization the gas distribution is expected to be highly inhomogeneous \cite{Pawlik2009, Shull2012, Finlator2012}. These inhomogenities, toghether with the spatial distribution and clustering properties of ionizing sources, imply that the ionization fraction is a spatially varying quantity at a given redshift. As  $\tau_{\rm es}$ depends on the column density of free electrons along the line of sight, the patchy nature of reionization generates fluctuations in the optical depth. Patchy reionization also produces secondary anisotropies in the CMB via the kinetic Sunyaev-Zel'dovich (KSZ) effect, related to the peculiar motion of ionized bubbles \cite{Sunyaev1980,Dvorkin2009}. Investigating the patchiness and morphology of the reionization process could provide crucial information on the astrophysical properties of the primeval galaxies, and on the distribution of ionized and neutral matter during the cosmic dawn \cite{2005MNRAS.360.1063S}. 

Future CMB experiments like the the Simons Observatory\footnote{https://simonsobservatory.org}, CMB Stage-IV\footnote{https://cmb-s4.org/} (CMB S4), and Probe of Inflation and Cosmic Origins\footnote{https://zzz.physics.umn.edu/ipsig/start} (PICO) will significantly surpass the \textit{Planck} mission as to sensitivity in temperature and polarization. These experiments have the potential to directly probe the patchiness in cosmic reionization by measuring non-Gaussian features imprinted on the CMB polarization spectra \cite{Dvorkin2009, Roy2018, Feng2018}. Complementary observations from radio-arrays like Hydrogen Epoch of Reionization Array\footnote{https://reionization.org/} (HERA) and Square Kilometer Array\footnote{https://www.skatelescope.org/} (SKA) operating in the MHz frequency range, will measure the 21cm angular distribution and power, so providing robust constraints on the distribution of HI gas out to very high redshifts. 

Since 21cm observations trace the evolution of neutral hydrogen HI and CMB observations of $\tau_{\rm es}$ trace the evolution of ionized hydrogen, it is expected that these probes are complementary to each other \cite{Meerburg2013, Ma2018, Alvarez2006, Alvarez2016}, and that their cross-correlation can greatly improve our knowledge of the cosmic reionization process. Cross-correlating $\tau_{\rm es}$ and 21cm observations can reduce the impact of systematic effects (e.g., related to foregrounds), and provide a tomographic mapping of the reionization process, allowed by the narrow frequency resolution of the 21cm radio arrays \cite{Meerburg2013}.

In the present paper we forecast the future detectability of the cross-correlation between fluctuations in $\tau_{\rm es}$ as probed by CMB experiments, and fluctuations in the 21cm differential brightness temperature $\Delta T_{\rm 21cm}$ at a given redshift as probed by radio interferometers. We also investigate the power of such $\tau_{\rm es}-21$cm cross-correlation in probing tomographically the patchiness of the cosmic reionization.

The plan of the paper is the following: in Section \ref{reiomodel} we describe our model of reionization; in Section \ref{xcorr} we define the observables entering our analysis, i.e., the 21cm differential brightness temperature $\Delta T_{\rm 21cm}$ and the optical depth for electron scattering $\tau_{\rm es}$, and provide the theoretical background for the $\tau_{\rm es}-21$cm cross correlation in terms of reionization morphology and of angular power spectrum; in Section \ref{reconstruction} we discuss the detectability of the $\tau_{\rm es}-21$cm cross correlation by future CMB experiments and radio arrays, and its dependence on various parameters describing reionization history and morphology; finally, in Section \ref{summary} we summarize our findings.

Throughout this work we adopt the \textit{Planck} 2018 \cite{Planck2018} cosmology with rounded parameters values: matter density $\Omega_M = 0.32$, dark energy density $\Omega_\Lambda=0.63$, baryon density $\Omega_b = 0.05$, Hubble constant $H_0 = 100\, h$ km s$^{-1}$ Mpc$^{-1}$ with $h = 0.67$, and mass variance $\sigma_8 = 0.81$ on a scale of $8\, h^{-1}$ Mpc. Primordial hydrogen and helium mass fractions $X_p = 0.75$ and $Y_p = 0.25$ are assumed.
Reported stellar masses and star formation rates (SFRs) or luminosities of galaxies refer to the Chabrier (2003) Initial Mass Function (IMF).

\section{Fiducial model of cosmic reionization}\label{reiomodel}

In this Section we describe our fiducial model for the cosmic reionization history. We envisage that faint, high-redshift star-forming galaxies are the primary source of ionizing photons. Thus the ionization rate is just proportional to the cosmic star formation history
\begin{equation}
\dot N_{\rm ion}\approx f_{\rm
esc}\, k_{\rm ion}\, \rho_{\rm SFR}\,;
\end{equation}
here $k_{\rm ion}\approx 4\times 10^{53}$ $(M_\odot/\rm yr)^{-1}$ is the number of ionizing photons per solar mass formed into stars, with the quoted value appropriate for the Chabrier IMF; $f_{\rm esc}\lesssim 10\%$ is the (poorly constrained) average escape fraction for ionizing photons from the interstellar medium of high-redshift galaxies \citep[see][]{Mao2007, Dunlop2013, Robertson2015, Lapi2017}; $\rho_{\rm SFR}(z)$ is the cosmic star formation density.

For $\rho_{\rm SFR}$ we follow the approach of \citet{Lapi2017} and \citet{Madau2014}, and compute it by integrating the observed luminosity functions from dust-corrected UV \citep[e.g., ][]{Bouwens2015, Bouwens2016}, far-IR \citep[e.g., ][]{Lapi2011, Gruppioni2013, Gruppioni2015, Rowan2016} and radio \citep[e.g., ][]{Novak2017} data down to a limiting UV magnitude $M_{\rm UV}^{\rm lim}$ considered to contribute to the ionizing background. Nowadays, the UV luminosity functions out to $z\lesssim 10$ are well determined down to a limit $M_{\rm UV}^{\rm lim}\lesssim -17$ from blank field surveys, and down to $M_{\rm UV}^{\rm lim}\approx -15$ when gravitational lensing by foreground galaxy clusters is exploited \citep[see][]{Alavi2014, Alavi2016, Livermore2017, Bouwens2016, Bouwens2019}. However, to efficiently reionize the Universe an extrapolation down to even fainter magnitudes, typically around $M_{\rm UV}^{\rm lim}\lesssim -13$ is necessary; note that due to the steepness in the faint end of the luminosity function, the resulting $\rho_{\rm SFR}$ is somewhat sensitive to the precise $M_{\rm UV}^{\rm lim}$ adopted. At $z\gtrsim 10$ the constraints on the UV luminosity functions are scanty, so the cosmic SFR density $\rho_{\rm SFR}$ has been extrapolated in that redshift range from the lower-$z$ behavior; a posteriori we have checked that the impact of this extrapolation is minor.

The competition between ionization and recombination determines the evolution of the fraction $x_{\rm HII}$ of ionized hydrogen via the equation \citep[see][]{Madau1999, Ferrara2014}:
\begin{equation}
\dot x_{\rm HII} = {\dot
N_{\rm ion}\over \bar n_{\rm
H}}-{x_{\rm HII}\over
t_{\rm rec}}\,,
\label{ionfrac}
\end{equation}
where $\bar n_{\rm H}\approx 2\times 10^{-7}$ cm$^{-3}$ is the mean comoving hydrogen number density. In addition, the recombination timescale reads $t_{\rm rec}\approx 2$ Gyr $[(1+z)/8]^{-3}\, C_{\rm HII}^{-1}$, where the case B coefficient for an IGM temperature of $2\times 10^4$ K has been used \cite{Lapi2017}; this timescale crucially depends on the clumping factor of the ionized hydrogen, for which a fiducial value $C_{\rm HII}\approx 3$ is usually adopted at the relevant redshifts \citep[see][]{Pawlik2013}.

Finally, the electron scattering optical depth $\tau_{\rm es}$ out to redshift $z$ is given by
\begin{equation}\label{eq_tau_eerm}
\tau_{\rm es}(<z) =c\,
\sigma_{\rm T}\, \bar n_{\rm
H}\,\int_0^{z}{\rm
d}z'\,{(1+z')^2\over H(z')}\,f_e\,(1+\delta_b)\,x_{\rm
HII}(z')\,,
\end{equation}
where $c$ is the speed of light, $\sigma_T$ is the Thomson scattering cross section, $H(z)=H_0\, [\Omega_M\,(1+z)^3+\Omega_\Lambda]^{1/2}$ the Hubble parameter, $f_e$ is the fraction of electrons per hydrogen nucleus (taking into account the presence of He), and $\delta_b$ the local baryon overdensity (usually neglected for the sky-averaged optical depth but important for related fluctuations, see Sect. \ref{sec|basic}). We compute the factor $f_e\simeq (1+Y_p/4\,X_p)\approx 1.083$ under the approximation of singly ionized He (see \cite{Kuhlen2012}), but note that its precise value has a negligible impact on the resulting $\tau_{\rm es}$ evolution (e.g., for doubly ionized He, $f_e\simeq 1+Y_p/2\,X_p\approx 1.167$).

\begin{figure}[t]
\begin{center}
\includegraphics[width=0.6\textwidth]{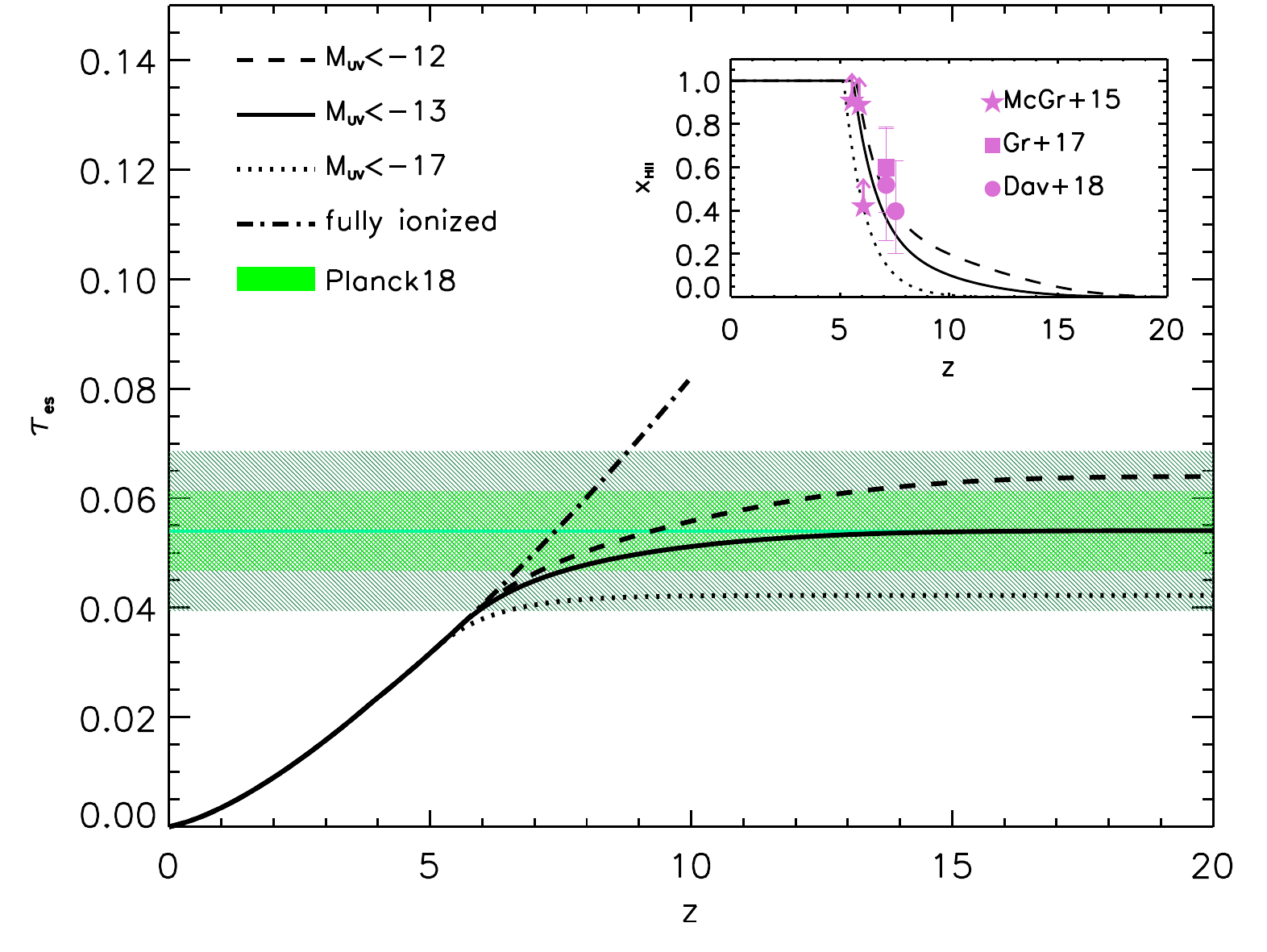}
\caption{Redshift evolution of the electron scattering optical depth $\tau_{\rm es}$. Thick dotted, solid, and dashed lines represent our fiducial model for the SFR density integrated down to UV-magnitude limits $M_{\rm UV}\approx -17$, $-13$, and $-12$, respectively. For reference, the dot-dashed line refers to a fully ionized universe up to redshift $z$. The green line shows the measurement (with the $1\sigma$ and $2\sigma$ uncertainty regions highlighted by the dark and light green areas respectively) from the \textit{Planck} collaboration 2018 \cite{Planck2018}. The inset illustrates the corresponding redshift evolution of the ionized hydrogen fraction $x_{\rm HII}$; constraints from the spectra of two highest-redshift quasars \cite{2017MNRAS.466.4239G, 2018ApJ...864..142D} and from the incidence of dark pixels in Ly$\alpha$ and Ly$\beta$ forests \cite{2015MNRAS.447..499M}.}
\label{tau}
\end{center}
\end{figure}

Figure~\ref{tau} shows the redshift evolution of the optical depth $\tau_{\rm es}$ (and of the corresponding ionized hydrogen fraction $x_{\rm HII}$ in the inset); this has been computed from our SFR density integrated down to different UV magnitude limits $M_{\rm UV}^{\rm lim}$, assuming a standard value $f_{\rm esc}\approx 5\%$ for the escape fraction of ionizing photons. For $M_{\rm UV}^{\rm lim}\approx -13$, the result (black solid line) agrees with the value of the optical depth for electron scattering $\tau_{\rm}\approx 0.054$ recently measured by the \textit{Planck} \cite{Planck2018}. This will constitute our fiducial reionization history in the $\tau_{\rm es}-$21 cross-correlation study. For reference, the dot-dashed line represents the optical depth expected in a fully ionized Universe up to redshift $z$; this is to show that the bulk of the reionization process occurred at $z \sim 6-8$ \citep[see][]{Schultz2014}. 

Adopting $M_{\rm UV}^{\rm lim}\approx -17$ corresponds to the observational limits of current blank-field UV surveys at $z\gtrsim 6$; the resulting optical depth (black dotted line) approaches the lower boundary of the 2$\sigma$ region allowed by \textit{Planck} data. At the other end, assuming $M_{\rm UV}^{\rm lim}\approx -12$ makes the resulting optical depth (black dashed line) to approach the upper boundary of the 2$\sigma$ region from \textit{Planck} measurements. The same upper and lower boundaries can be also obtained by retaining the UV limiting magnitude $M_{\rm UV}^{\rm lim}\approx -13$ but varying the escape fraction $f_{\rm esc}$ from the fiducial value of $5\%$ to around $10\%$ and $2\%$, respectively.
Actually the degeneracies among the parameters entering the computation of the optical depth can be highlighted via the expression 
$f_{\rm esc}\, k_{\rm ion}\, C_{\rm HII}^{−0.3}\, \Gamma[\alpha+2, 10^{-0.4\,(M_{\rm UV}^{\rm lim})-M_{\rm UV}^*}]\approx$ const, where $\Gamma$ is the incomplete gamma function, $\alpha$ is the faint-end slope of the SFR function, and $M_{\rm UV}^*\sim -21$ is the UV luminosity beyond which the SFR functions features an exponential fall-off; this comes out just by representing the SFR function through a Schechter functional shape with redshift dependent parameters, see \cite{Lapi2017} and  Fig. 6 in \cite{Lapi2015} for further details). For example, switching from a Chabrier to a Salpeter IMF would imply fewer ionizing photons per unit SFR, so a reduction in the parameter $k_{\rm ion}$ of a factor $\sim 1.6$; to obtain the reionization history corresponding to the \textit{Planck} best-fit value would then require increasing $f_{\rm esc}$ by the same amount, or extending the UV limiting magnitude from $M_{\rm UV}^{\rm lim}\sim -13$ down to $-11.5$. 

In the present paper we do not attempt to constrain these parameters via the $\tau_{\rm es}-21$cm cross-correlation; rather we have set their plausible ranges of values just by requiring the resulting reionization histories to be consistent, within the uncertainties, with the \textit{Planck} measurements of the optical depth. We will discuss in Section \ref{reconstruction} how the $\tau_{\rm es}-21$cm cross-correlation analysis is affected when switching among such reionization histories, and hence how it depends on the above parameters, and especially on the most uncertain $M_{\rm UV}^{\rm lim}$ and $f_{\rm esc}$.

In the way of comparing with previous works, it is worth noticing that Meerburg et al. (2013; \cite{Meerburg2013}) have conducted a similar study of the $\tau_{\rm es}-$21 cross correlation basing on a phenomenological representation of the cosmic reionization history; specifically, they implemented a `tanh' shape of the ionization fraction (see their Fig. 1) with parameters set to produce a value of the optical depth around $\tau_{\rm es}\approx 0.084$. As a matter of fact, their reionization history features a quite steep growth of the ionization fraction around redshift $11$, with reionization being almost completed  (ionization fraction exceeds $80\%$) around redshift $z\approx 10$. Our reionization histories are different in two respects: being gauged on the \textit{Planck} 2018 best fit value of $\tau_{\rm es}\approx 0.054$, reionization is shifted toward lower redshift (ionization fraction is $50\%$ at redshift $z\approx 7$); being based on realistic evolution of the ionizing power from the observed galaxy luminosity function, the ionization history is more gradual than in a tanh model. We stress that the main differences in our and their results concerning the cross power-spectra and the detectability forecasts are mainly related to such diverse shapes in the adopted reionization histories.

\section{$\tau_{\rm es}-$21cm cross correlation}\label{xcorr}

In this Section we discuss how patchy reionization generates fluctuations in the optical depth as well as in the 21cm differential brightness temperature.

\subsection{Basic quantities: $\Delta T_{\rm 21cm}$ and $\delta\tau$}\label{sec|basic}
The optical depth $\tau_{\rm 21cm}$ of the neutral hydrogen medium to the hyperfine transition is given by \citep{Field1958}
\begin{equation}
\tau_{\rm 21cm}=\frac{3c^3 \hslash\,A_{10}}{16 k\,\nu_{\rm 21cm}^2\, T_S}\,\frac{x_{\rm HI}\, \bar n_{\rm H}\,(1+z)^3}{(1+z)\,{\rm d}v_\parallel/{\rm d r_\parallel}}\,.
\end{equation}
Here $\nu_{\rm 21cm}\approx 1420$ MHz is the rest-frame frequency of the 21cm hyperfine transition line, $A_{10}\approx 2.85\times 10^{-15}$ s$^{-1}$ is the spontaneous emission coefficient, $T_S$ is the spin temperature which regulates the intensity of the radiation, $x_{\rm HI}$ is the neutral hydrogen fraction, and $v_\parallel$ is the proper velocity along the line of sight (l.o.s.); at high redshift peculiar motions along the l.o.s. are small compared to the Hubble flow and $(1+z)\,{\rm d}v_\parallel/{\rm d r_\parallel}\simeq H(z)$ holds to a very good approximation.

The differential 21cm brightness temperature $\Delta T_{\rm 21cm}$ is the difference between the redshifted 21cm brightness temperature and the redshifted CMB sky-averaged temperature $T_{\rm CMB}(z)\approx 2.73\,(1+z)$ K;
for small $\tau_{\rm 21cm}$, it just reads \citep{Furlanetto2006}
\begin{equation}\label{deltatb}
\Delta T_{\rm 21cm}\simeq  \frac{T_S-T_{\rm CMB}}{1+z}\,\tau_{\rm 21cm}\,.
\end{equation}
In the redshift range relevant for reionization where dark energy and radiation are negligible so that $H(z)\simeq H_0\, \Omega_M^{1/2}\, (1+z)^{3/2}$, Eq. (\ref{deltatb}) can be recast into the form \cite{Meerburg2013}
\begin{equation}\label{eq|21cm}
\Delta T_{\rm 21cm}\approx 23~{\rm mK}\, \left(\frac{1+z}{8}\right)^{1/2}\,(1+\delta_b)\,x_{\rm HI}\,\left[\frac{T_S-T_{\rm CMB}}{T_S}\right]\,.
\end{equation}

During the reionization process, the ionized fraction $x_{\rm HII}(z,\uvec{n})\simeq \bar x_{\rm HII}\,(1+\delta x_{\rm HII})$ is expected to depend on l.o.s. direction $\uvec{n}$, and thus to fluctuate with respect to the sky-averaged value $\bar x_{\rm HII}(z)$; together with the presence of a local baryon overdensity $(1+\delta_b)$ this will induce fluctuations $\delta\tau_{\rm es}$ in the optical depth for electron scattering $\tau_{\rm es}$, which is indeed proportional to the integral of $(1+\delta_b)\,x_{\rm HII}$ along the l.o.s. after Eq. (\ref{eq_tau_eerm}). On the other hand, the complementary neutral hydrogen fraction $x_{\rm HI}=1-x_{\rm HII}$ will also vary along the l.o.s. direction, so inducing (together with $\delta_b$) fluctuations $\delta(\Delta T_{\rm 21cm})$ in the 21cm differential brightness temperature $\Delta T_{\rm 21cm}\propto (1+\delta_b)\,x_{\rm HI}$. Plainly, such fluctuations $\delta(\Delta T_{\rm 21cm})$ and $\delta\tau_{\rm es}$ can be connected to each other; for $T_S> T_{\rm CMB}$ and the redshift range relevant to reionization, the relation writes down as \citep{Holder2006, Meerburg2013}
\begin{equation}\label{eq|dtau21}
\delta \tau_{\rm es}=0.0035\,\int {\rm d}z\,\left[(1+z)^{\frac{1}{2}}\,\delta_b-\frac{\delta[\Delta T_{\rm 21cm}(z)]}{8.8 \, \rm mK}\right]\,.
\end{equation}
This is routinely exploited to build the halo model of the $\tau_{\rm es}-21$cm cross-correlation presented below, that we will in turn use to forecast the detectability of the signal.

\subsection{Morphology of reionization}\label{sec:reio_morphology}

In order to express the fluctuations $\delta\tau_{\rm es}$ and  $\delta (\Delta T_{\rm 21cm})$ in a dimensionless way, we introduce the brightness temperature field $\Psi$ and ionization fraction field $X$:
\begin{equation}
\Psi(z,\uvec{n})=\frac{T_S-T_{\rm CMB}}{T_S}\,(1+\delta_b)\,x_{\rm HI}\,,
\end{equation}

\begin{equation}
X(z,\uvec{n})=(1+\delta_b)\,x_{\rm HII}\,.
\end{equation}

We assume the standard picture envisaging that ionizing sources (e.g., galaxies in our framework) start to ionize the surrounding regions via closely spherical bubbles. As time passes, ionized bubbles progressively overlap and then merge with each other, inducing eventually a complete reionization of the Universe. 
Semi-analytic modeling and numerical simulations focused on the morphology of cosmic reionization \cite{Zahn2007,Zahn2011,Friedric2011,Majumdar2014,Lin2016,Ronconi2020} indicate a log-normal distribution of the bubble sizes, in the form
\begin{equation}
P(R)=\frac{1}{R}\frac{1}{\sqrt{2\pi\sigma_{\rm lnr}^2}}\exp{\left[-\frac{\{\ln\left(R/\bar{R}\right)\}^2}{2\sigma_{\rm lnr}^2}\right]}\,;
\end{equation}
for future reference, typical values for the mean bubble radius $\bar{R}\approx 5$ Mpc and for the $1\sigma$ dispersion $\sigma_{\rm lnr}\approx \log(2)$ apply around the reionization redshift $z\sim 7$.

The three-dimensional power spectrum of the cross correlation between the fluctuation field $\delta\Psi$ and $\delta X$  can be decomposed into two terms:
\begin{equation}\label{ptotal_eq}
P_{\delta X\delta\Psi}(k)= P^{1b}_{\delta X\delta\Psi}(k)+P^{2b}_{\delta X\delta\Psi}(k)\,,
\end{equation}
where $P^{1b}_{\delta X\delta\Psi}$ is the 1-bubble contribution coming from the distribution of neutral and ionized regions inside an individual bubble, and $P^{2b}_{\delta X\delta\Psi}$ is the 2-bubble contribution from regions located in different bubbles. 

The 1-bubble term can be expressed as \cite{Wang2006}
\begin{equation}\label{p1b_eq}
P^{1b}_{\delta X\delta\Psi}(k)= -x_{\rm HII}(1-x_{\rm HII})\left[\alpha (k)+\beta(k)\right],
\end{equation}
where the quantity $\alpha(k)$ and $\beta({k})$ are given by 
\begin{equation}\label{alphak_eq}
\alpha(k)=\frac{\int {\rm d}R\, P(R)\, [V(R)\,W(KR)]^2}{\int {\rm d}R\, P(R)\,V(R)},
\end{equation}
\begin{equation}\label{betak_eq}
\beta(k)=\int \frac{{\rm d}^3\bold{k}'}{(2\pi)^3}\,P\left(\left | \bold{k}-\bold{k}' \right |\right)\,\alpha(k')\,,
\end{equation}
and involve the matter power spectrum $P(k)$, the volume of $V(R)=\frac{4}{3}\pi R^3$ of a bubble with size $R$, and a filtering window function (Fourier transform of a top-hat in real space) expressed as
\begin{equation}
W(kR)=3\,(kR)^{-3}\,\left[\sin(kR)-kR\,\cos(kR)\right]\,.
\end{equation}
We adopt the approximation from \cite{Wang2006} and calculate $\beta(k)$ as
\begin{equation}
\beta({k})=\frac{P(k)\, \sigma_R^2\, \int {\rm d}R\, P(R )\,V(R)}{\{P^2(k)+[\sigma_R^2 \,\int {\rm d}R\, P(R)\, V(R)]^2]^{1/2}}\,,
\end{equation}
in terms of the mass variance $\sigma_R$ of the matter power spectrum filtered on the scale $R$.

In addition, the 2-bubble term is given by
\begin{equation}\label{p2b_eq}
P^{2b}_{\delta X\delta \Psi}(k)= \left[(1-x_{\rm HII})\ln(1-x_{\rm HII})\gamma(k)-x_{\rm HII}\right]^2\,P(k),
\end{equation}
where $\gamma(k)$ is defined as
\begin{equation}\label{gamma_eq}
\gamma(k)= b\times \frac{\int {\rm d}R\, P(R)\, V(R)\, W(kR)}{\int {\rm d}R\, P(R)\, V(R)}\,,
\end{equation}
and involves the clustering bias $b$ of the ionized bubbles with respect to the spatially-average matter density. Hereafter we consider a linear bias with a fixed value $b\approx 6$ \cite{Wang2006}.

\subsection{Angular Power Spectrum}

We now translate the above three-dimensional expressions into an angular power spectrum on the sky
by the usual multipole expansions via spherical harmonics $Y_{\ell m}$. 

For the 21cm brightness temperature fluctuations the harmonic coefficients are given by 
\begin{equation}\label{alm21}
a_{\ell m}^{\rm 21cm}=4\pi (-i)^{\ell}\int \frac{{\rm d}^3k}{(2\pi)^3}\,\delta\Psi(k)\,I_{\ell}^{\rm 21cm}(k)\,Y^\star_{\ell m}
\end{equation}
where
\begin{equation}\label{Ilm21}
I_{\ell}^{\rm 21cm}(k)=T_0(z)\,\int_0^\infty {\rm d}\chi\, W(z,\chi)\,J_{\ell}(k\chi)\,;
\end{equation}
here $T_0(z)\approx 23\,[(1+z)/8]^{1/2}$ mK is the redshift-dependent prefactor in Eq. (\ref{eq|21cm}), $\chi$ is the comoving distance to redshift $z$, and $W(z,\chi)$ is an observational Gaussian band filter centered at $z$. The latter accounts for the fact that any ground-based radio array has a narrow frequency resolution, that in turn determines a resolution in redshift or in comoving distance
\begin{equation}\label{dchi}
\Delta\chi\approx \left(\frac{\Delta \nu}{0.1 \rm MHz}\right)\,\left(\frac{1+z}{10}\right)^{1/2}\,;
\end{equation}
this is relevant in empowering a tomographic analysis of the cross-correlation signal.
In the Limber approximation the auto-power spectrum is written then as
\begin{equation}
C_\ell^{2121}\simeq T_0^2(z)\,\int \frac{{\rm d}\chi}{\chi^2}\,W^2(z,\chi)\,P_{\delta\Psi\delta\Psi}\left(\chi, k=\frac{\ell}{\chi}\right)\,,
\end{equation}
where $P_{\delta\Psi\delta\Psi}$ is the auto power spectrum of the 21cm fluctuations $\delta \Psi$.

For the optical depth fluctuation field the harmonic coefficients are given by
\begin{equation}
a_{\ell m}^{\tau}=4\pi (-i)^{\ell}\, \int \frac{{\rm d}^3k}{(2\pi)^3}\,\delta X(k)\,I_{\ell}^{\tau}(k)\, Y^\star_{\ell m}
\end{equation}
where
\begin{equation}
I_{\ell}^{\tau}(k)=\bar n_{\rm H}\sigma_T\, f_e\int \frac{{\rm d}\chi}{a^2}\, J_{\ell}(k\chi)\,.
\end{equation}
The related angular auto power spectrum can be written as:
\begin{equation}
C_\ell^{\tau\tau}\simeq \sigma^2_T\bar n_{\rm H}^2\, f_e^2\, \int \frac{{\rm d}\chi}{a^4\chi^2}\,P_{\delta X\delta X}\left( \chi, k=\frac{\ell}{\chi}\right)\,,
\end{equation}
where $P_{\delta X\delta X}$ is the power spectra of the ionized hydrogen fraction fluctuations $\delta X$.

\begin{figure}[t]
\begin{center}
\includegraphics[width=0.6\textwidth]{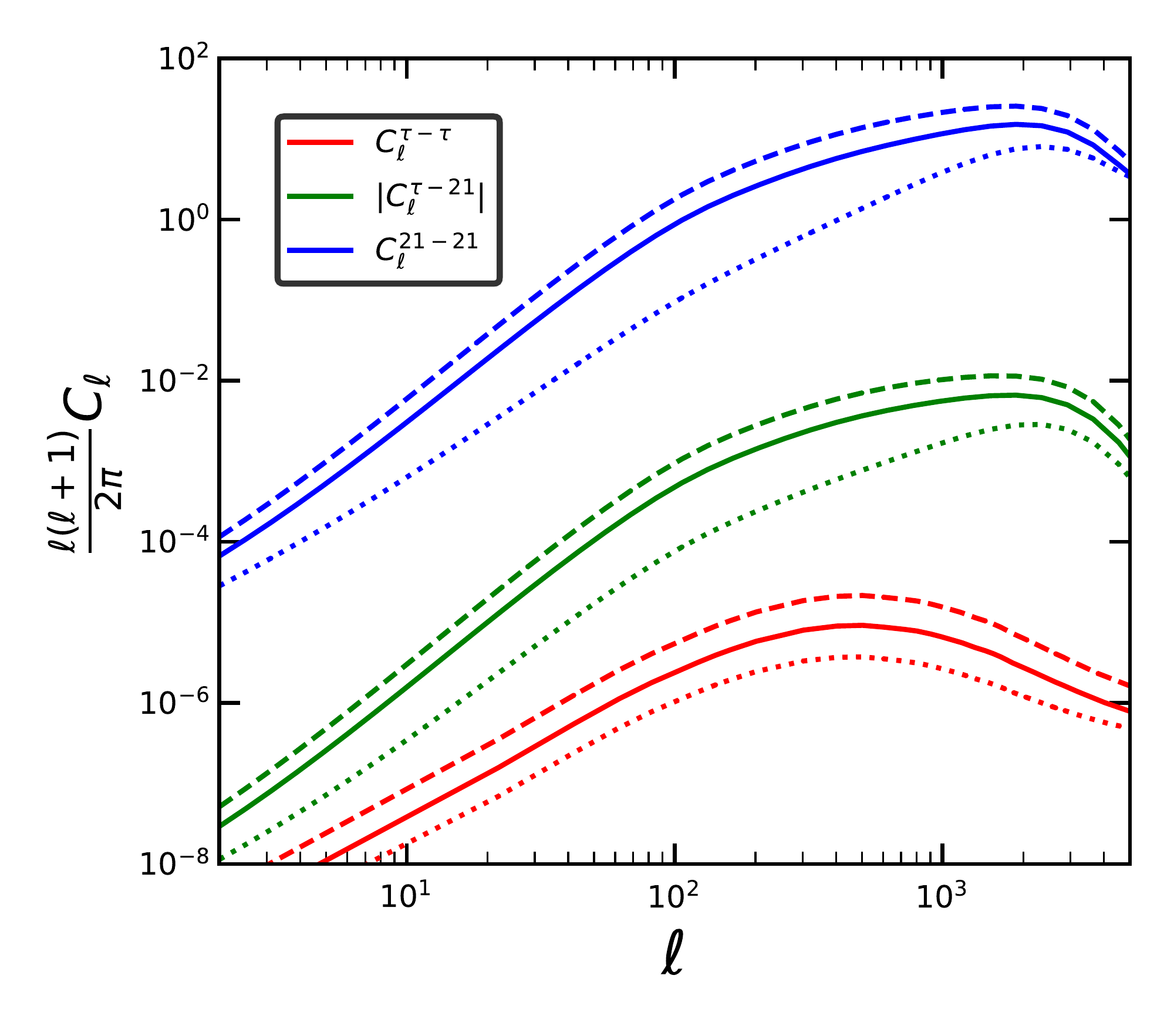}
\caption{Auto and cross power spectra of the $\tau_{\rm es}$ and 21cm fluctuations, by adopting the three reionization histories of Section \ref{reiomodel} that correspond to spatially-averaged optical depths $\tau\approx 0.054$ (solid lines) and $\tau\approx 0.046$ (dotted lines) and $\tau\approx 0.070$ (dashed lines) bracketing the \textit{Planck} measurements; the bubble size distribution parameters $\bar R=5$ Mpc and $\sigma_{\rm lnr}=\log(2)$ are adopted, see the text for details.}
\label{allspec}
\end{center}
\end{figure}

Finally, the cross-correlation angular power spectrum reads
\begin{equation}
\left<a_{\ell m}^\tau a_{\ell m}^{21*}\right>=\int \frac{dk}{k}\left[\frac{k^3P_{X\Psi}}{2\pi^2}\right]I_\ell^\tau(k)I_{\ell}^{\rm 21cm}(k)
\end{equation}
and in terms of the usual $C_\ell$ coefficients can be written as \cite{Meerburg2013}
\begin{equation}
C_{\ell}^{\tau 21}\simeq T_0(z)\,\bar n_{\rm H}\sigma_T\, f_e\int_0^\infty \frac{{\rm d} \chi}{a^2\, \chi^2}\, W(z,\chi)\,P_{\delta X \delta\Psi}\left(\chi,k=\frac{\ell}{\chi}\right)\,.
\end{equation}
We show in Figure \ref{allspec} the auto and angular power spectra $C_\ell^{\tau\tau}$, $C_\ell^{2121}$ and $|C_\ell^{\tau21}|$ at a redshift $z\approx 7$ for the three reionization histories of Section \ref{reiomodel} that bracket the \textit{Planck} determination of the spatially-averaged optical depth; as suggested by semi-analytic work and numerical simulations \cite{Zahn2007,Zahn2011,Friedric2011,Majumdar2014,Lin2016,Ronconi2020}, we have adopted a reference log-normal bubble size distribution with parameters $\bar R=5$ Mpc and $\sigma_{\rm lnr}=\log(2)$ as appropriate around the reionization redshift $z\approx 7$ (see also Sect. \ref{sec:reio_morphology}). In absolute values, the $21-$cm auto-correlation spectrum is the largest and the $\tau_{\rm es}$ auto-correlation spectrum is the smallest, while the $\tau_{\rm es}-21$cm cross-correlation spectrum strikes an intermediate course; this reflects the relative smallness of fluctuations $\delta\tau_{\rm es}$ in optical depth with respect to those $\delta(\Delta T_{\rm 21 cm})$ in differential brightness temperature after Eq. (\ref{eq|dtau21}). Reionization histories featuring a higher integrated value of $\tau_{\rm es}$ yield generally larger auto and cross-correlation spectra, since they corresponds to larger fluctuations in ionized and neutral hydrogen fractions.

\begin{figure}[h!]
\begin{center}
\includegraphics[width=0.6\textwidth]{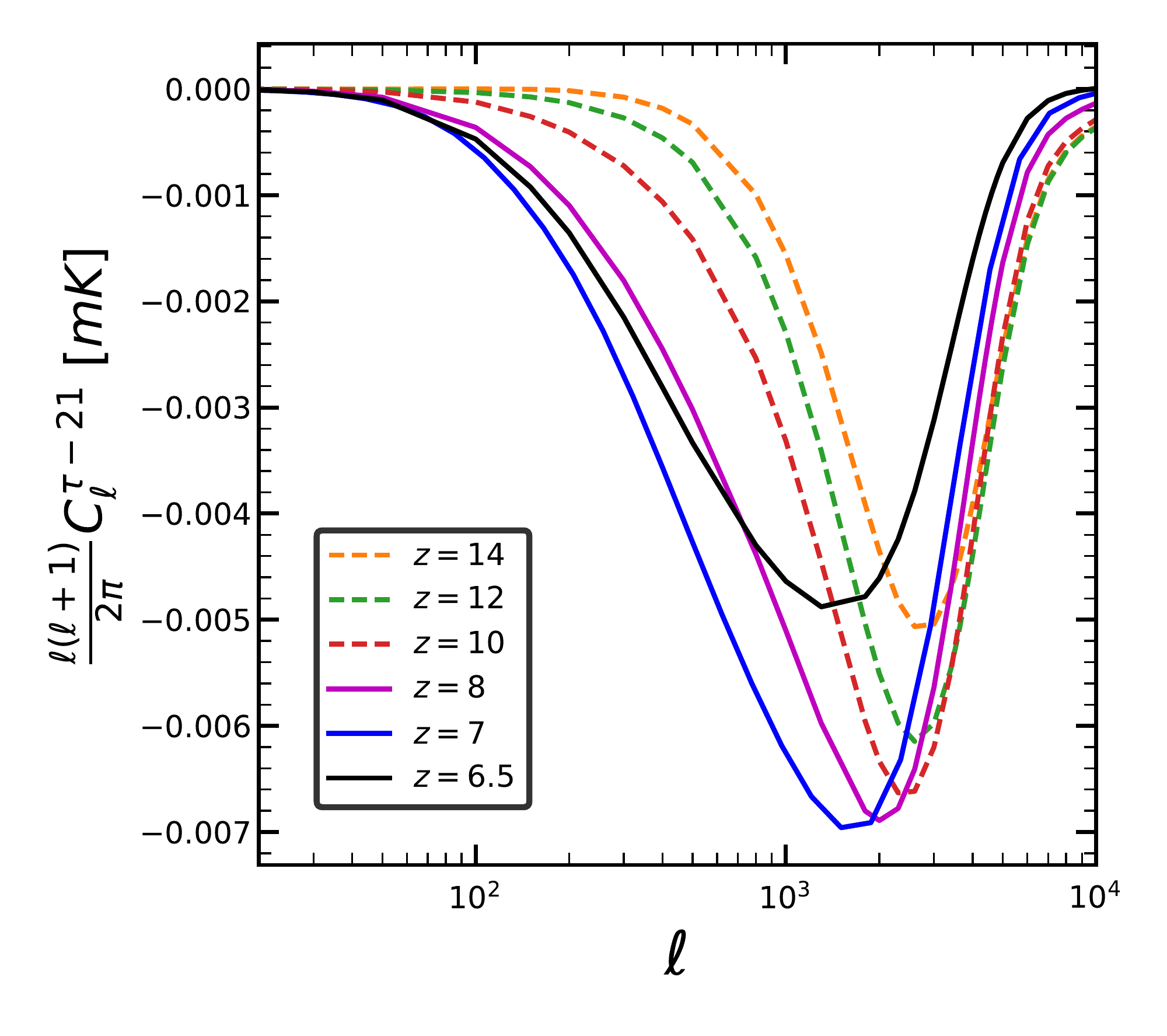}
\caption{Redshift evolution of the cross power spectrum $C_\ell^{\tau-21}$ down and around to the reionization redshift $z\approx 7$, for our fiducial reionization history of Section \ref{reiomodel} corresponding to $\tau=0.054$ and parameters of the ionized bubble distribution as in previous Figure.}
\label{spec_redshift_evol}
\end{center}
\end{figure}

In Figure \ref{spec_redshift_evol}, the $\tau_{\rm es}-21$cm cross spectrum is illustrated as a function of redshift, for the same parameters reported above. The cross correlation signal is negative, featuring an inverse bell shape with a minimum at around the multipole corresponding to the average size of the ionizing bubbles, and a width mirroring that of the bubble size distribution. The depth of the minimum is maximal at the redshift where the ionizing fraction is around $50\%$ and then becomes shallower in moving at lower and at higher redshift; the cross power spectrum vanishes in a completely neutral or completely ionized Universe. These two Figures show that precise measurements of the cross-spectrum would yield a detailed view on the reionization history of the Universe (see the pioneering work by \cite{2005MNRAS.360.1063S}).

\section{Detectability of $\tau_{\rm es}-21$cm cross correlation}\label{reconstruction}

We now turn to discuss the detectability of the $\tau_{\rm es}-21$cm cross correlation. 

\subsection{Noise model}
The uncertainty of the angular cross power spectrum can be calculated as
\begin{equation}
(\Delta C_{\ell}^{\tau 21})^2=\frac{1}{(2\ell+1)f_{\rm sky}}\left[(C_{\ell}^{\tau 21})^2+(C_\ell^{\tau\tau}+N_{\ell}^{\tau\tau})(C_\ell^{2121}+N_{\ell}^{2121})\right]\,;
\end{equation}
where $f_{\rm sky}$ is the observed sky fraction, $C_\ell^{\tau \tau}$ and $C_\ell^{2121}$ are the optical depth and 21cm brightness temperature fluctuation auto power spectra, corresponding to the reionization history and morphology described in previous Sections, and $N_\ell^{\tau \tau}$ and $N_\ell^{2121}$ are the corresponding noise power spectra.

Here we summarize the reconstruction method of $\tau_{\rm es}$, which is basically performed by applying an estimator to the polarized CMB. CMB quadrupole radiation is altered via Thomson scattering by the inhomogeneous distribution of free electrons generated during patchy reionization. The temperature and polarization pattern of the CMB is thus modulated as $T(\uvec{n})=T_0(\uvec{n})+\int \delta\tau T_1(\uvec{n})$ and $(Q\pm iU)(\uvec{n})=(Q\pm iU)_0(\uvec{n})+\int \delta\tau (Q\pm iU)_1(\uvec{n})$, where $T_0$ and $(Q\pm iU)_0$ are the temperature and polarization parameters at the recombination whereas $T_1$ and $(Q\pm iU)_1$ are the response fields due to patchy reionization. In addition to that, primary CMB temperature and polarization fluctuations are screened by a factor $e^{\tau\pm \delta{\tau(\uvec{n})}}$ due to patchy reionization. 

Patchy reionization also introduces a correlation among different Fourier modes, which can be expressed in terms of a coupling factor $f^{\tau}$ such that
\begin{equation}
\langle X(\boldsymbol{\ell_1})X'(\boldsymbol{\ell_2})\rangle=(2\pi)^2C_\ell^{XX'}\delta(\boldsymbol{L})+f^{\tau}_{XX'}(\boldsymbol{\ell_1},\boldsymbol{\ell_1})\delta \tau(\boldsymbol{L})\,;
\end{equation}
here $X$ and $X'$ could be any combination of $T$,$E$ and $B$, while  $\boldsymbol{L}=\boldsymbol{\ell_1}+\boldsymbol{\ell_2}$. In principle, one can apply minimum variance quadratic estimator to any combination of $X$ and $X^{\prime}$ and reconstruct the $\tau_{\rm es}$ field. In the present context, we only consider the EB estimator as it provides the highest signal-to-noise ratio (SNR). In the flat sky approximation, the coupling factor reads $f^{\tau}_{EB}(\boldsymbol{\ell_1},\boldsymbol{\ell_1})=(\tilde{C}_{\ell_1}^{EE}-\tilde{C}_{\ell_1}^{BB})\sin2(\phi_{\ell_1}-\phi_{\ell_2})$, where $\tilde{C}_{\ell_1}^{EE}$ and $\tilde{C}_{\ell_1}^{BB}$ are the $E$-mode and $B$-mode power spectra including the effect of patchy reionization and $\phi_\ell=\cos^{-1}(\uvec{n}\dot{\ell})$. The expectation value of the estimator becomes
\begin{equation}
    \langle \hat{\tau}_{EB}(\boldsymbol{\ell_1})\hat{\tau}_{EB}(\boldsymbol{\ell_2})\rangle = (2\pi)^2\delta(\boldsymbol{\ell_1},\boldsymbol{\ell_2})\left[C_L^{\tau\tau}+\tilde{N}^\tau_{EB}(\boldsymbol{L})\right]\,,
\end{equation}
in terms of the reconstruction noise $\tilde{N}^\tau_{EB}(\boldsymbol{L})$. 

Following \citet{Dvorkin2009}, we compute $N_\ell^{\tau\tau}$ in the form
\begin{equation}
    \tilde{N}_{EB}^{\tau}(\boldsymbol{L})=\left[\int\frac{d^2\ell_1}{(2\pi)^2}f^\tau_{EB}(\boldsymbol{\ell_1}, \boldsymbol{\ell_2})F_{EB}^{\tau}(\boldsymbol{\ell_1},\boldsymbol{\ell_2})\right]^{-1}\,;
\end{equation}
here $F_{EB}^{\tau}$ acts like a filter which optimizes the variance of the estimator and can be written as
\begin{equation}
    F_{EB}^{\tau}(\boldsymbol{\ell_1},\boldsymbol{\ell_2})=\frac{f_{EB}^{\tau}(\boldsymbol{\ell_1},\boldsymbol{\ell_2})}{(\tilde{C}_{\ell_1}^{EE}+N_{\ell_1}^{P})(\tilde{C}_{\ell_2}^{BB}+N_{\ell_2}^{P})}\,.
\end{equation}

This term depends on both the reionization history and morphology, as well as on the instrumental noise of a given CMB experiment. The latter reads $N_\ell^{P}=\Delta^2_P \exp\left[\frac{\ell (\ell+1)\Theta^2_f}{8\ln(2)} \right]$, where $\Delta_P$ is the noise of the polarization detector in units of $\mu K$-arcmin (which is $\sqrt{2}$ times bigger than the detector noise for temperature), and $\Theta_f$ is the FWHM of the beam. Setting the instrumental noise and $\ell_{\rm max}=3000$, we compute $N_\ell^{\tau\tau}$.
\begin{table}[t]
\begin{center}
    \begin{tabular}{| c | c | c | c | c |}
    \hline
    CMB Experiment & Sensitivity $\Delta_T$ & $\theta_f$ \\
     & [$\mu$K arcminute]& [arcminute] \\ \hline
    CMB-S4  &1 & 1 \\ \hline
    PICO  & 0.6 & 2 \\ \hline
    Simons Observatory  & 3 & 2 \\
    \hline
    \end{tabular}
    \caption{Configurations of CMB experiments considered in our analysis (see text).}
    \label{table_cmbex}
\end{center}
\end{table}
\begin{table}[h!]
\begin{center}
    \begin{tabular}{| c | c | c | c | c | c | c |}
    \hline
    21cm Experiment & $A_{\rm eff}$ & $t_{\rm int}$ & $\Delta \nu$ \\
      & [$m^2$] & [hours] & [$Mhz$] \\ \hline
    HERA 350 & 53878 & 1080 & 0.1 \\ \hline
    SKA  & 416596 & 1080 & 0.2\\ \hline
    \end{tabular}
    \caption{Configurations of 21cm observatories considered in our analysis (see text).}
   \label{table_21cmex}
\end{center}
\end{table}

$N_\ell^{2121}$ should include both the radiometer noise and a noise contribution from 21cm foreground. Since the 21cm foreground noise is poorly understood, we will ignore it in this Section, so obtaining optimistic estimates of the SNR. We will then introduce a simple foreground model in Section \ref{forg} and discuss its effects on the detectability of the $\tau_{\rm es}-21$cm cross-correlation signal.

21cm thermal noise angular power spectra is expected to be smooth and it is given by \cite{2004ApJ...608..622Z}
\begin{equation}
    \frac{\ell(\ell+1)}{2\pi}N_\ell^{2121}=\frac{T^2_{sys}S^2_{\rm sky}}{\Delta\nu t_{\rm int}A^2_{\rm eff}}\frac{\ell(\ell+1)}{\ell^2_{\rm max}}\,;
\end{equation}
here $t_{\rm int}$ is the total integration time for the 21cm observation, $\Delta \nu$ is the bandwidth of the experiment, and $\ell_{\rm max}$ is the achievable maximum multipole. The effective area covered by antennae is $A_{\rm eff}$ and $S_{\rm sky}$ is the total area of the observed sky. $T_{sys}$ is the temperature of the system which accounts both for the antenna temperature ($T_{\rm ant}$) and the average sky temperature ($T_{\rm sky}$) due to the foreground contamination. For simplicity, we consider a constant antenna temperature of $40$ K and approximate the sky temperature as $T_{\rm sky}=5.0\,(\nu/710\,\rm M Hz)^{-2.6}$ K; we also adopt $S_{\rm sky}=5\times 10^6\,m^2$ \cite{2010ARA&A..48..127M,2013PhRvD..87f4026H}. Finally, when forecasting the SNR for the detectability of the $\tau_{\rm es}-21$cm cross-correlation signal, we also assume this noise is fixed even if the observed sky fraction $f_{\rm sky}$ is increased.

\subsection{Signal to noise ratio (SNR)}
The SNR for the detectability of the $C_\ell^{\tau-21}$ cross correlation signal is computed, in cumulative terms, as \cite{Tashiro2010}
\begin{equation}
\left(\frac{S}{N}\right)_z^2=f_{\rm sky}\,\sum_{\ell_{\rm min}}^{\ell_{\rm max}}(2\ell+1)\int_z dz' \frac{\left|C_{\ell}^{\tau 21}(z')\right|^2}{(C_\ell^{\tau\tau}+N_{\ell}^{\tau\tau})(C_{\ell}^{2121}(z')+N_{\ell}^{2121}(z'))}\,,
\end{equation}
at the redshift $z$ where the $C_\ell^{\tau-21}$ is probed, that is determined by the observational frequency of the 21cm experiment. 

The detectability of the $\tau_{\rm es}-21$cm power spectrum is investigated for the experiments listed in Table \ref{table_cmbex} (CMB) and Table \ref{table_21cmex} (radio arrays). All the following plots refer to redshift $z\sim 7$ and to the reference combination of CMB-S4$\times$SKA experiments.

\begin{figure*}[t]
    \centering
    \begin{subfigure}[t]{0.5\textwidth}
        \centering
         \includegraphics[height=8cm]{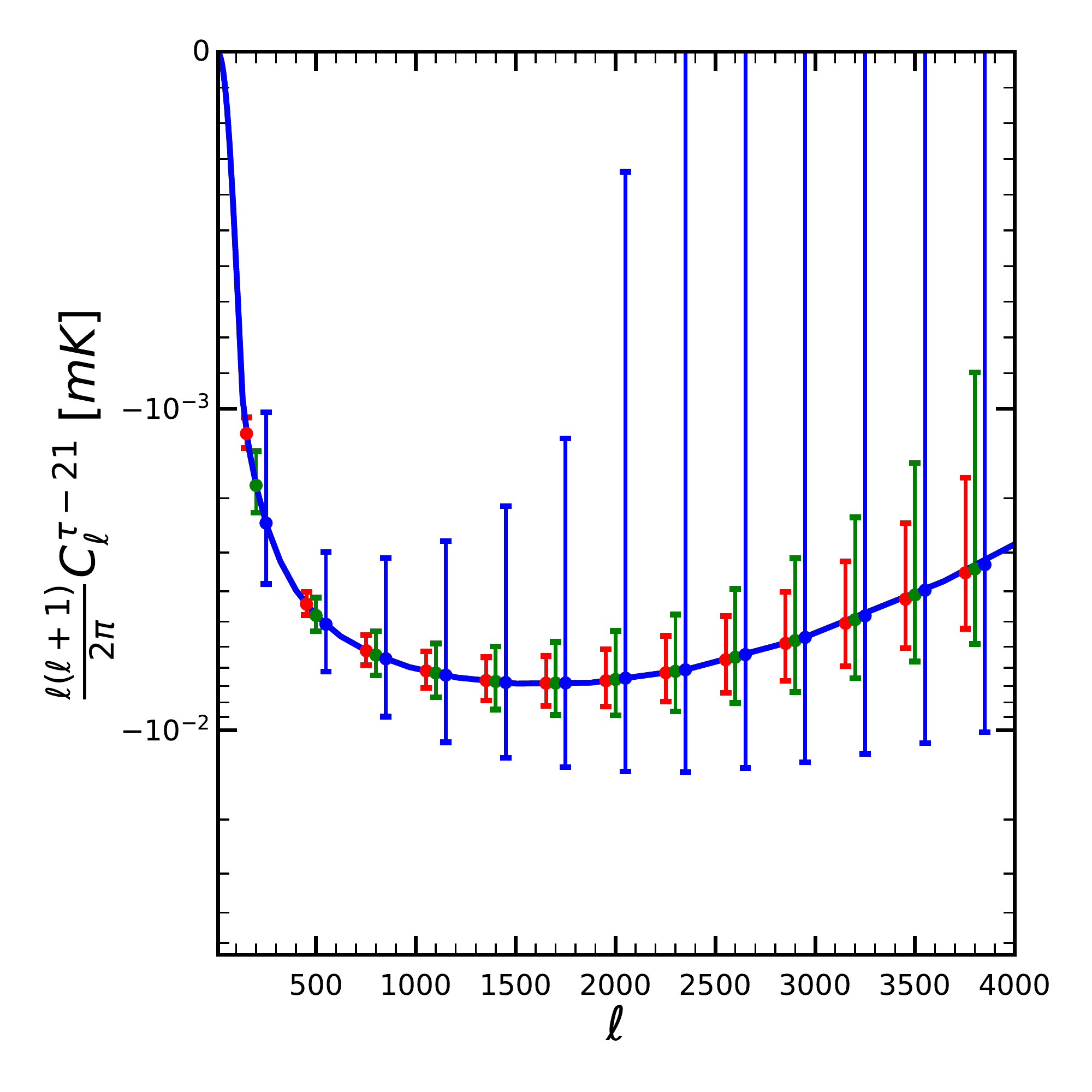}
    \end{subfigure}%
    ~
    \begin{subfigure}[t]{0.5\textwidth}
        \centering
    
        \includegraphics[height=8cm]{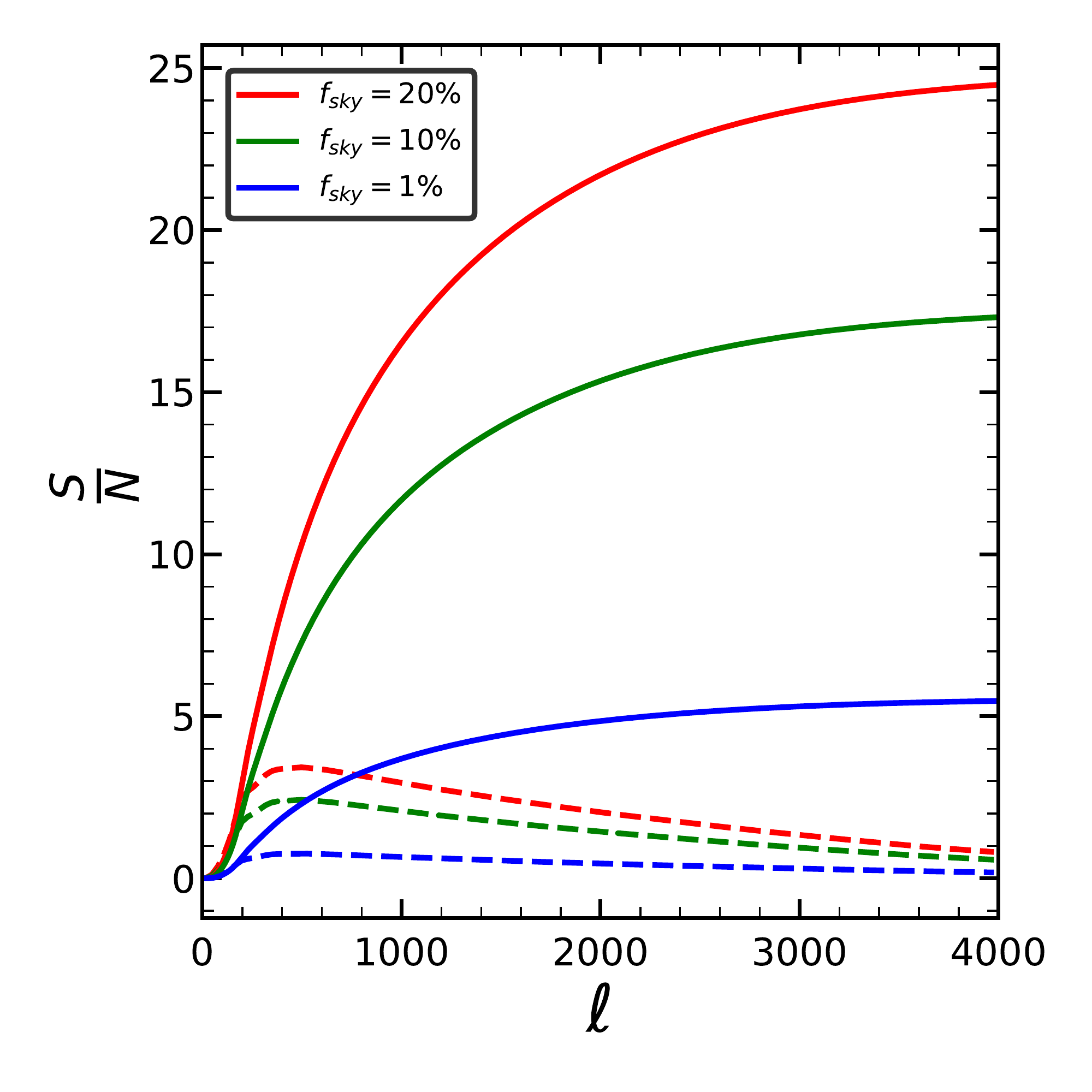}
    \end{subfigure}
    \caption{Dependence of the forecasted $C_\ell^{\tau-21}$ (left panel) and related SNR (right panel) on the observed sky fraction,  as labeled in the legend. In the right panel, solid lines refer to cumulative SNR while dashed lines to SNR in the binned spectra with $\Delta_\ell\approx 100$. The experiments considered here are CMB S4 and SKA.}
\label{cl-fsky}
\end{figure*}

In Figure \ref{cl-fsky} we show the detectability of the cross correlation signal as a function of the observed sky fraction. While future CMB experiments will observe more than $40\%$ of the sky, 21cm observations will be limited to a smaller portion of it ($\approx 1-10\%$). The SNR of the cross correlation increases as $\sqrt{f_{\rm sky}}$ in terms of the common sky fraction, so it is basically limited by the 21cm observations, and amounts to SNR $\sim 5.5-24.6$ for $f_{\rm sky}\sim 1\%-20\%$. Even with a sky fraction of a few percent, the global cross-correlation signal is detectable at more than $5\sigma$; for sky fractions larger than 10\% it will be possible to pick up the signal in the multipole range around the minimum of the cross-correlation at more than $3\sigma$.

\begin{figure*}[h]
    \centering
    \begin{subfigure}[t]{0.5\textwidth}
        \centering
         \includegraphics[height=8cm]{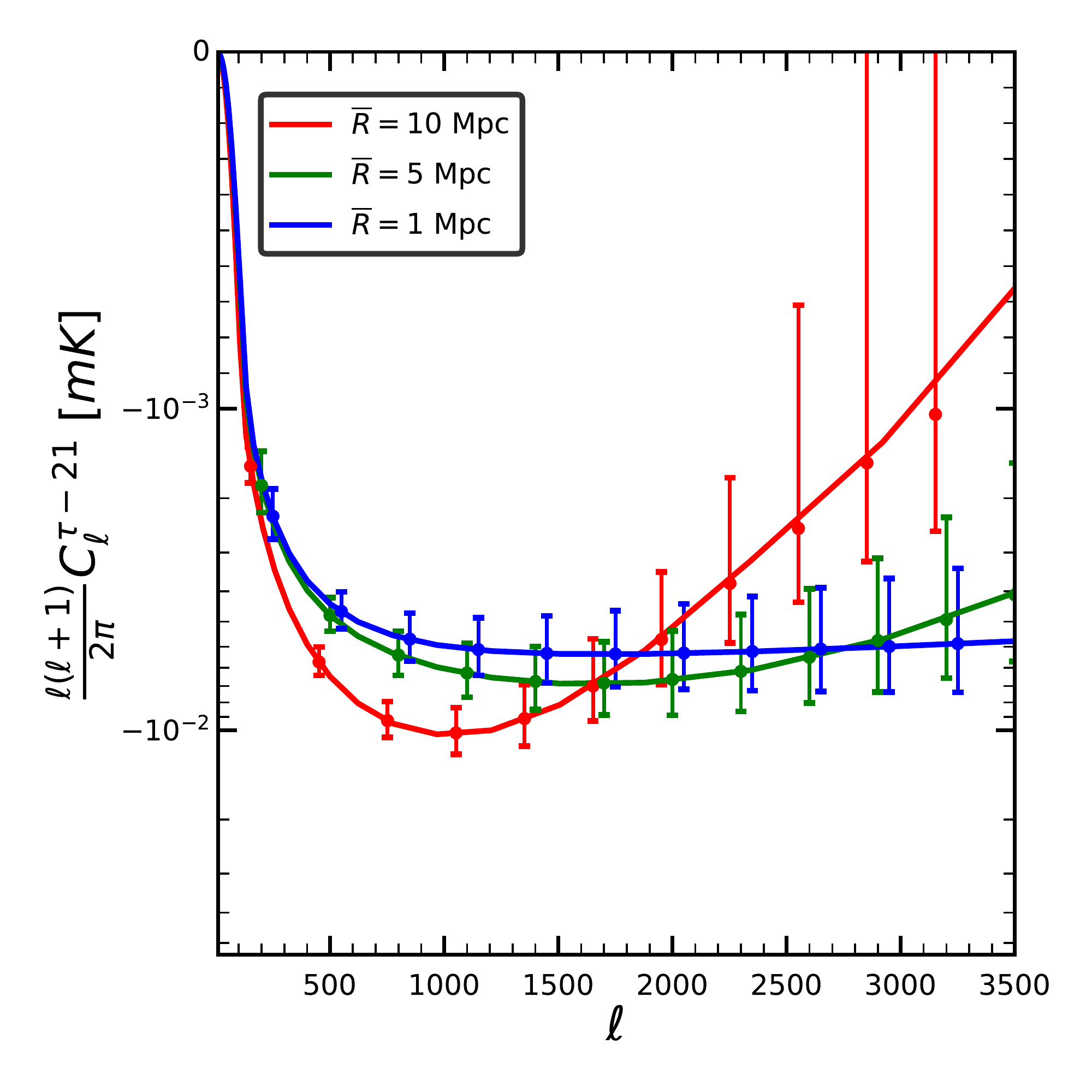}
    \end{subfigure}%
    ~
    \begin{subfigure}[t]{0.5\textwidth}
        \centering
    
       \includegraphics[height=8cm]{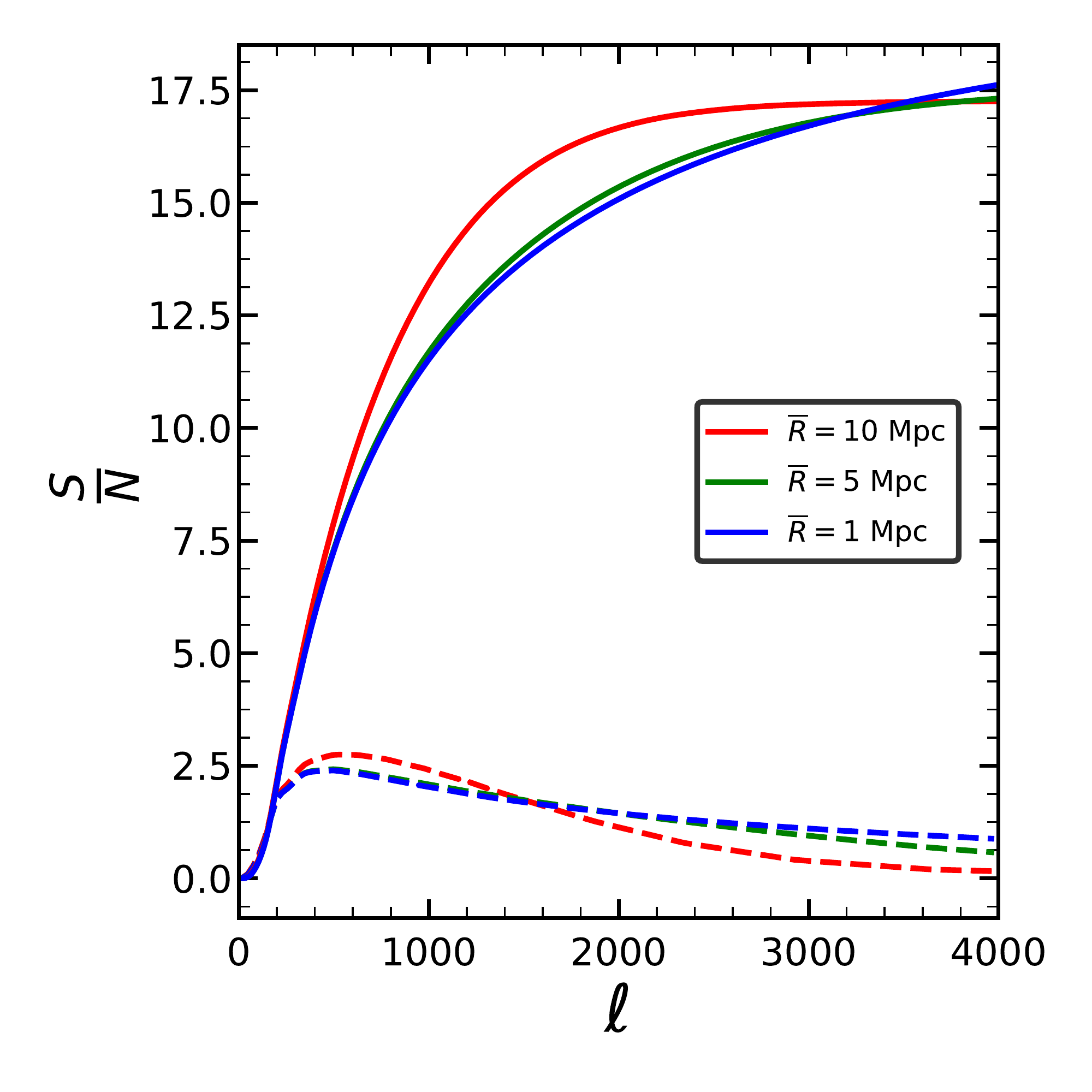}
    \end{subfigure}
     \caption{Same as previous figure for the average bubble size, as labeled in the legend.}
\label{cl-rbar}
\end{figure*}

\begin{figure}[h!]
\begin{center}
   \centering
    \begin{subfigure}[t]{0.5\textwidth}
       \centering
        \includegraphics[width=8cm]{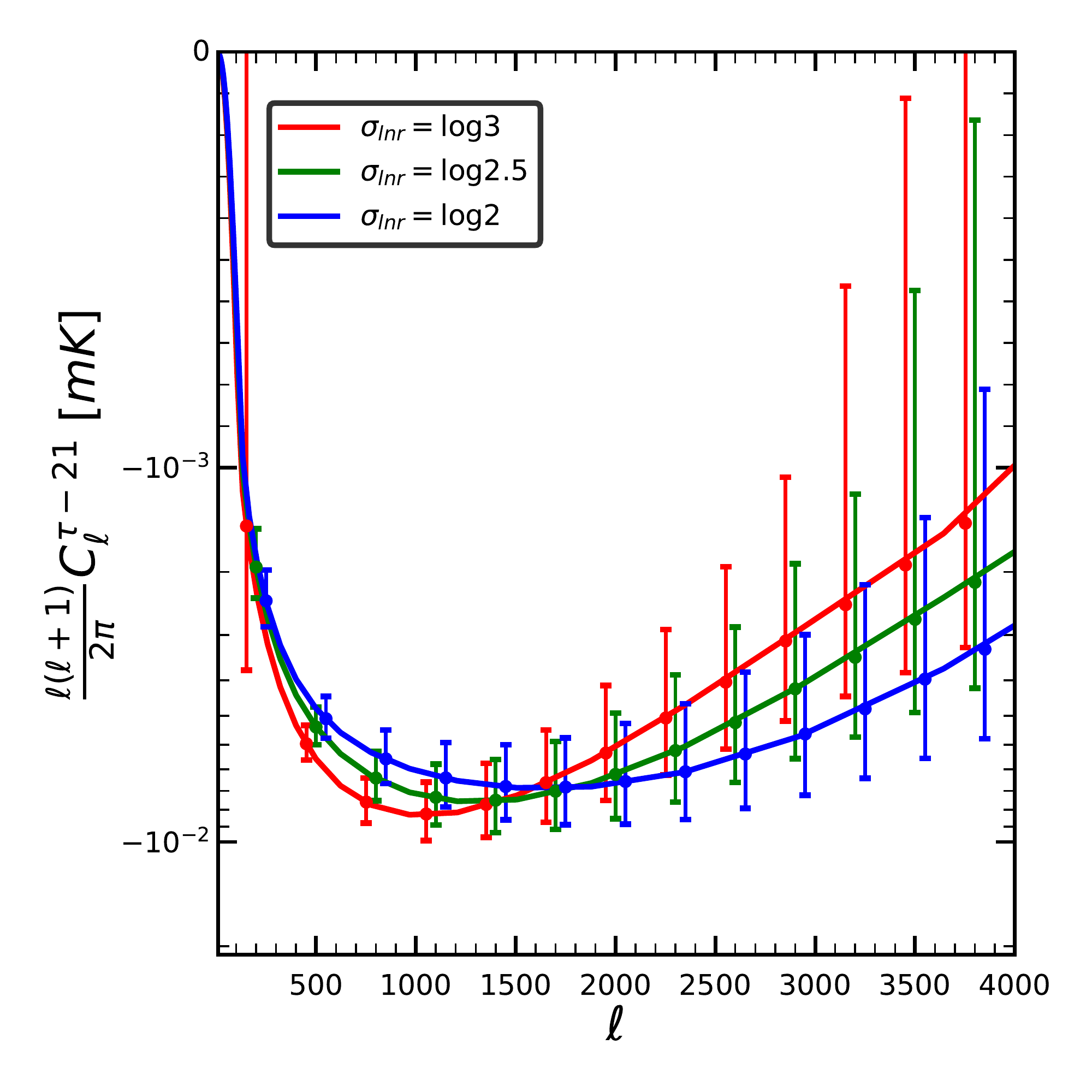}
    \end{subfigure}%
    ~
    \begin{subfigure}[t]{0.5\textwidth}
        \centering
\includegraphics[width=8cm]{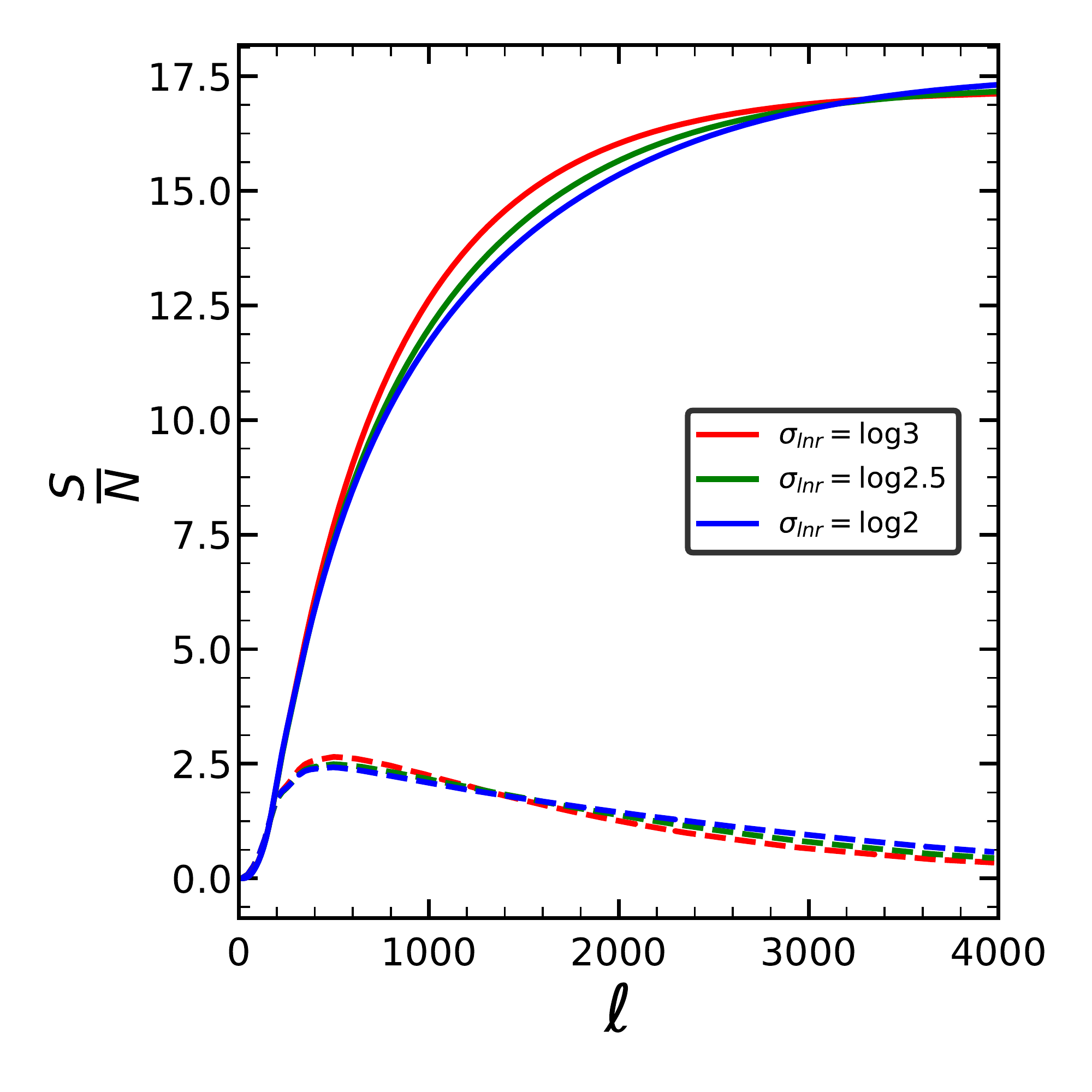}
    \end{subfigure}
    \caption{Same as previous figure for the dispersion of the bubble size distribution, as labeled in the legend.}
    \label{cl-sigma}
\end{center}
\end{figure}

In Figure \ref{cl-rbar} and \ref{cl-sigma} we illustrate how the detectability of the $\tau_{\rm es}-21$cm cross-correlation depends on the parameters describing the reionization morphology, i.e the mean bubble size $\bar{R}$ and the variance of its (log-normal) distribution $\sigma_{\rm lnr}$. Figure \ref{cl-rbar} shows that the cumulative SNR is marginally affected by those parameters; e.g., the SNR changes by $\sim 3\%$ when $\bar{R}$ increases from 1 to 10 Mpc. However, Figure \ref{cl-sigma} illustrates that the distribution of bubbles imprints clear signatures in the cross power spectrum $\ell (\ell+1)\, C_\ell^{\tau21}$; as already mentioned, the position of the minimum occurs at a multipole corresponding to the average bubble size, and its extent scales proportionally to the width of the bubble distribution. From this point of view, precision measurements of the cross power spectrum can pose intriguing constraints on the morphology of the reionization history.

\begin{figure}[h!]
\begin{center}
   \centering
    \begin{subfigure}[t]{0.5\textwidth}
        \centering
\includegraphics[width=8cm]{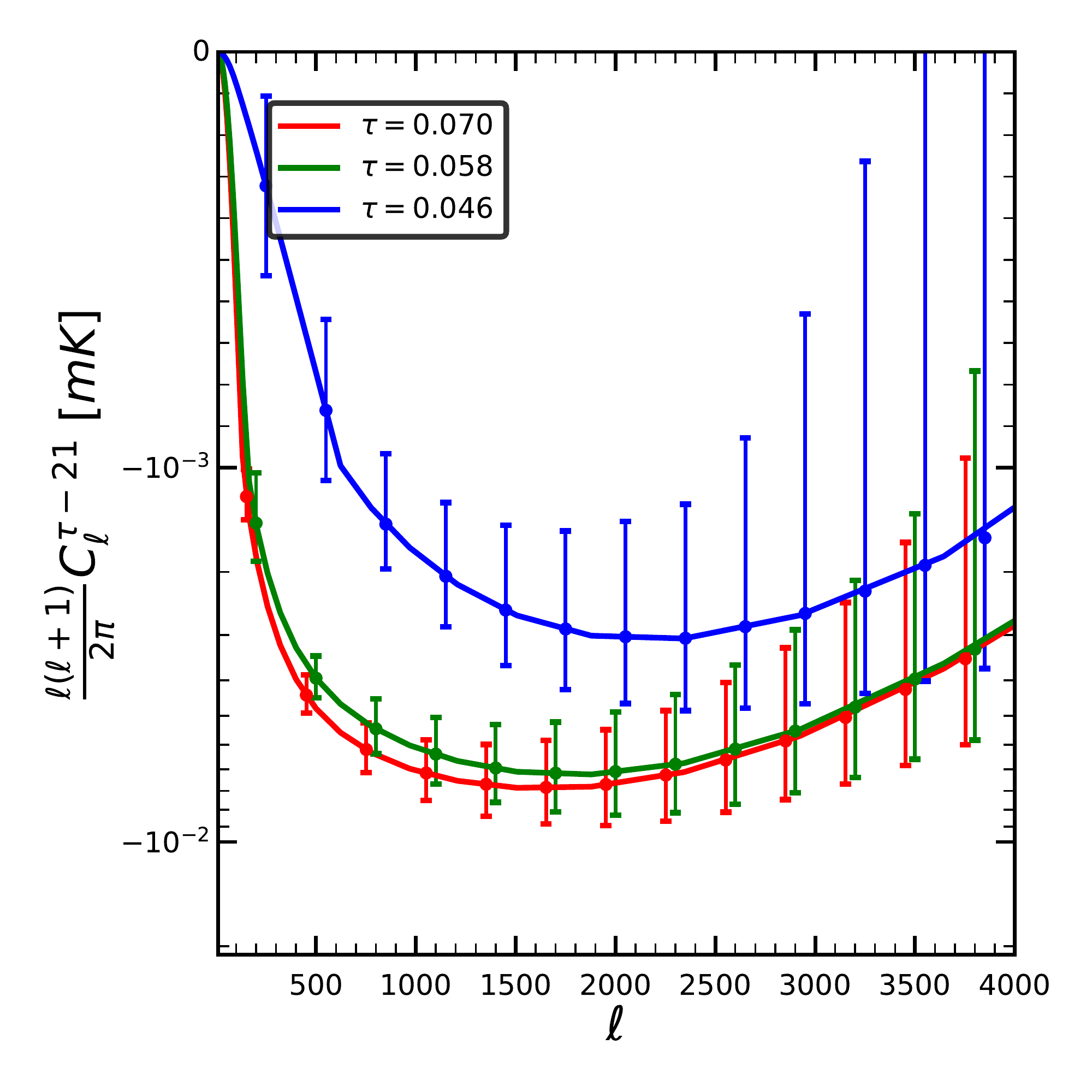}\\
    \end{subfigure}%
    ~
    \begin{subfigure}[t]{0.5\textwidth}
        \centering
\includegraphics[width=8cm]{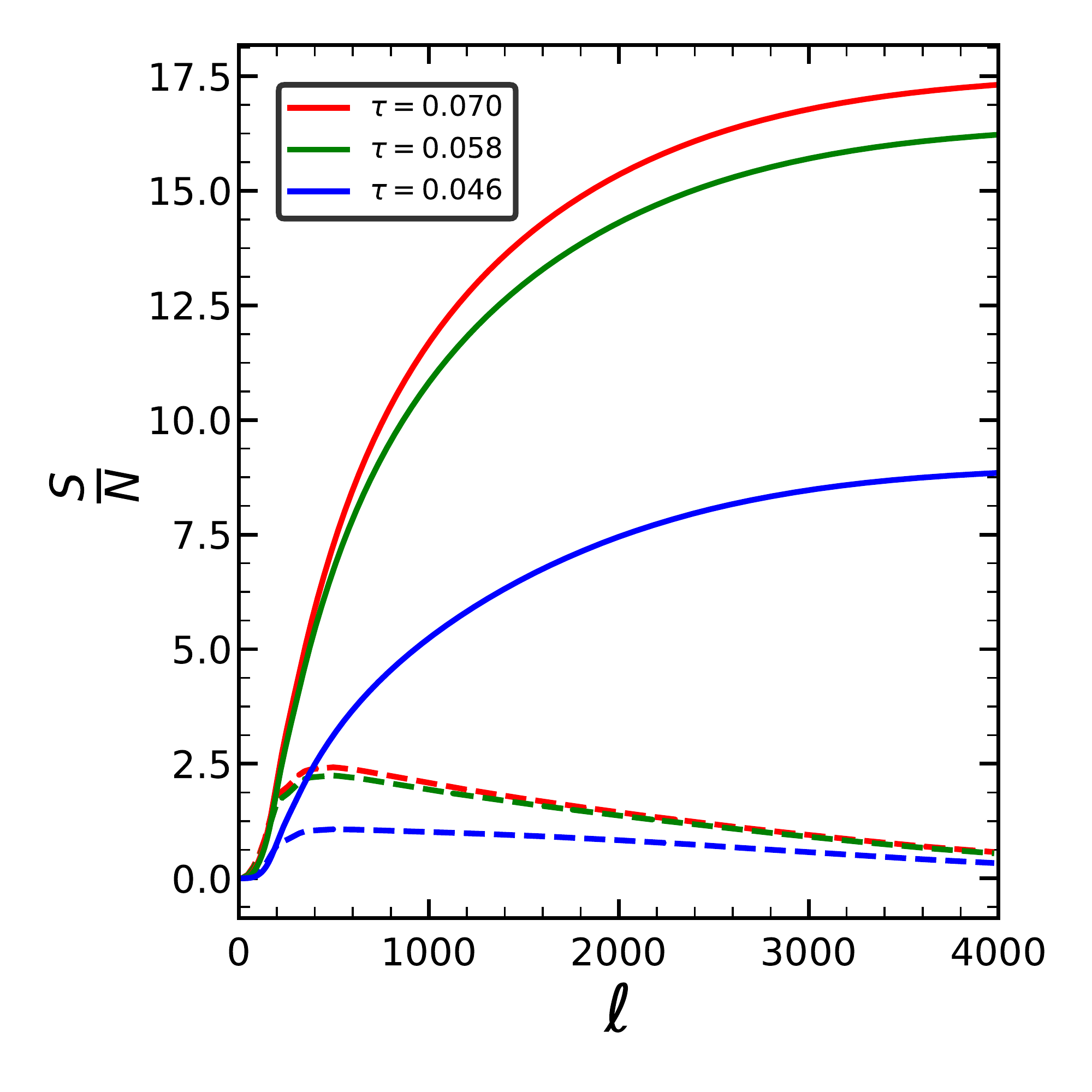}
    \end{subfigure}
    \caption{Same as previous Figure for the three reionization histories of Section \ref{reiomodel} corresponding to different spatially-averaged optical depth $\tau_{\rm es}$ as labeled in the legend.}
    \label{cl-tau}
\end{center}
\end{figure}

In Figure \ref{cl-tau} we investigate how the detectability of the cross-correlation is affected by assuming the three reionization histories of Section \ref{reiomodel}, corresponding to different values of the spatially-averaged optical depth for electron scattering bracketing the \textit{Planck} measurements (i.e $\tau=0.070, 0.58, \mathrm{and}\,\, 0.046$). In this particular case, we kept the morphology of reionization fixed by setting up $\sigma_{\rm lnr}=\log2 $ and $\bar{R}= 5$ Mpc. The cross power spectrum and cumulative SNR are slightly modified for $\tau> 0.05$ while they drastically changes for $\tau<0.05$ since for such low values of the optical depth the reionization process is far from being completed at the redshift $z\approx 7$ plotted here.

\begin{figure}[h!]
\begin{center}
\includegraphics[width=0.6\textwidth]{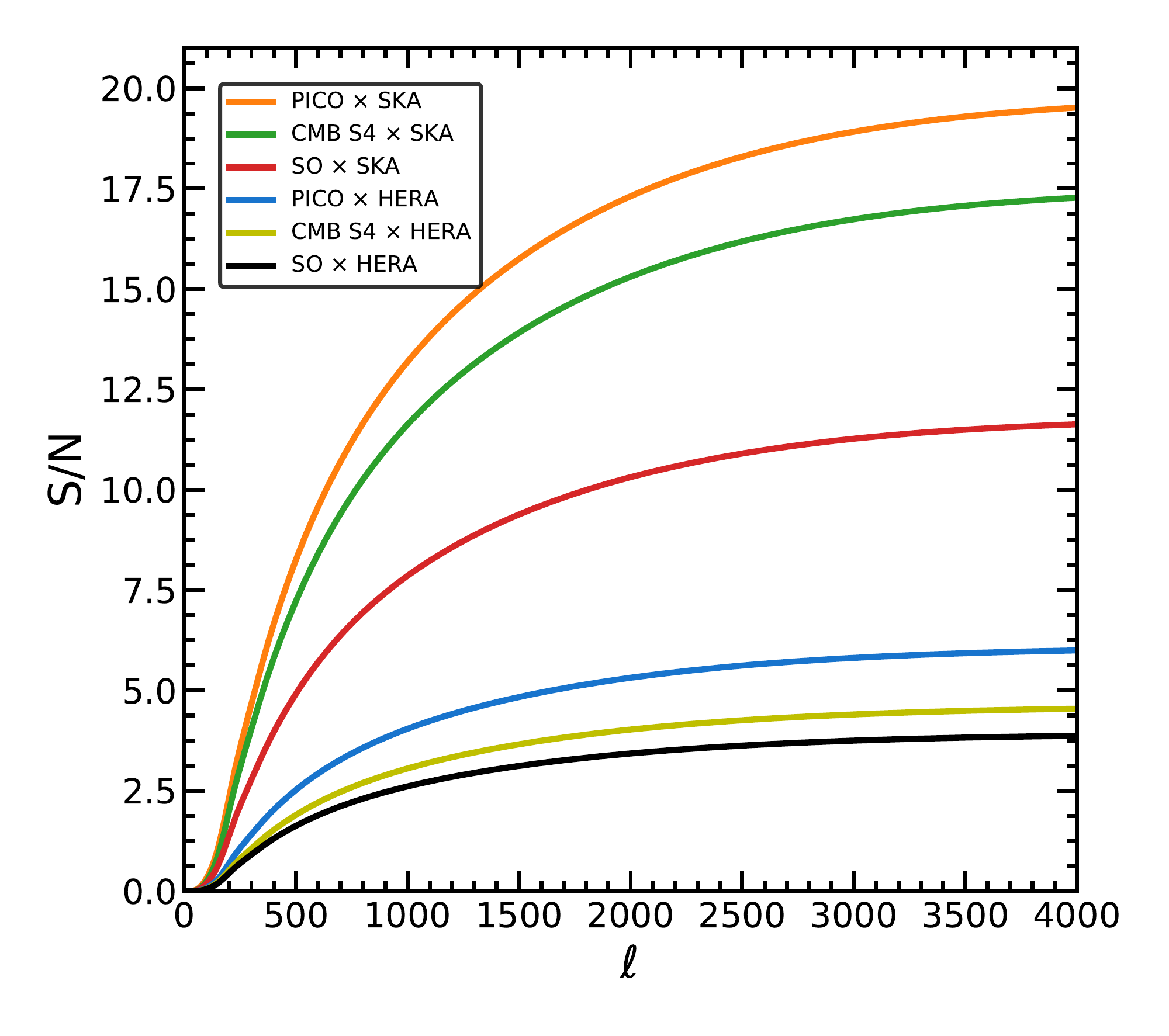}
\caption{Forecasted cumulative SNR of the $\tau_{\rm es}-21$cm cross correlation by combining different CMB and 21cm experiments, listed in Tables \ref{table_cmbex} and \ref{table_21cmex}.}
\label{SNR_ex_comb}
\end{center}
\end{figure}

In Figure \ref{SNR_ex_comb} we forecast the detectability of the $\tau_{\rm es}-21$cm cross-correlation for different combination of future CMB (CMB-S4, PICO, Simon Observatory) and 21cm experiments (HERA, SKA), whose features are listed in Table \ref{table_cmbex} and \ref{table_21cmex}. The computed SNR are well above $5\sigma$ for almost all the combinations, with the best sensitivity achieved from PICO$\times$SKA. 

\begin{table}[h!]
\begin{center}
    \begin{tabular}{| c | c | c | c | c | c | c |}
    \hline
    21cm Experiment  & $S/N$ \\ \hline
    CMB S4 $\times$ SKA & 17\\ \hline
    CMB S4 $\times$ HERA & 4.5\\ \hline
    
    PICO $\times$ SKA & 19.5\\ \hline
    PICO $\times$ HERA & 5.9\\ \hline
    
    SO $\times$ SKA & 11.5\\ \hline
    SO $\times$ HERA & 3.8\\ 
    \hline
    \end{tabular}

    \caption{Forecasted cumulative SNR of the $\tau_{\rm es}-21$cm cross correlation by combining different CMB and 21cm experiments.}
\end{center}
\label{table_sn}
\end{table}

\subsection{Effects of 21cm foregrounds}\label{forg}
\begin{figure}[h!]
\begin{center}
\includegraphics[width=0.6\textwidth]{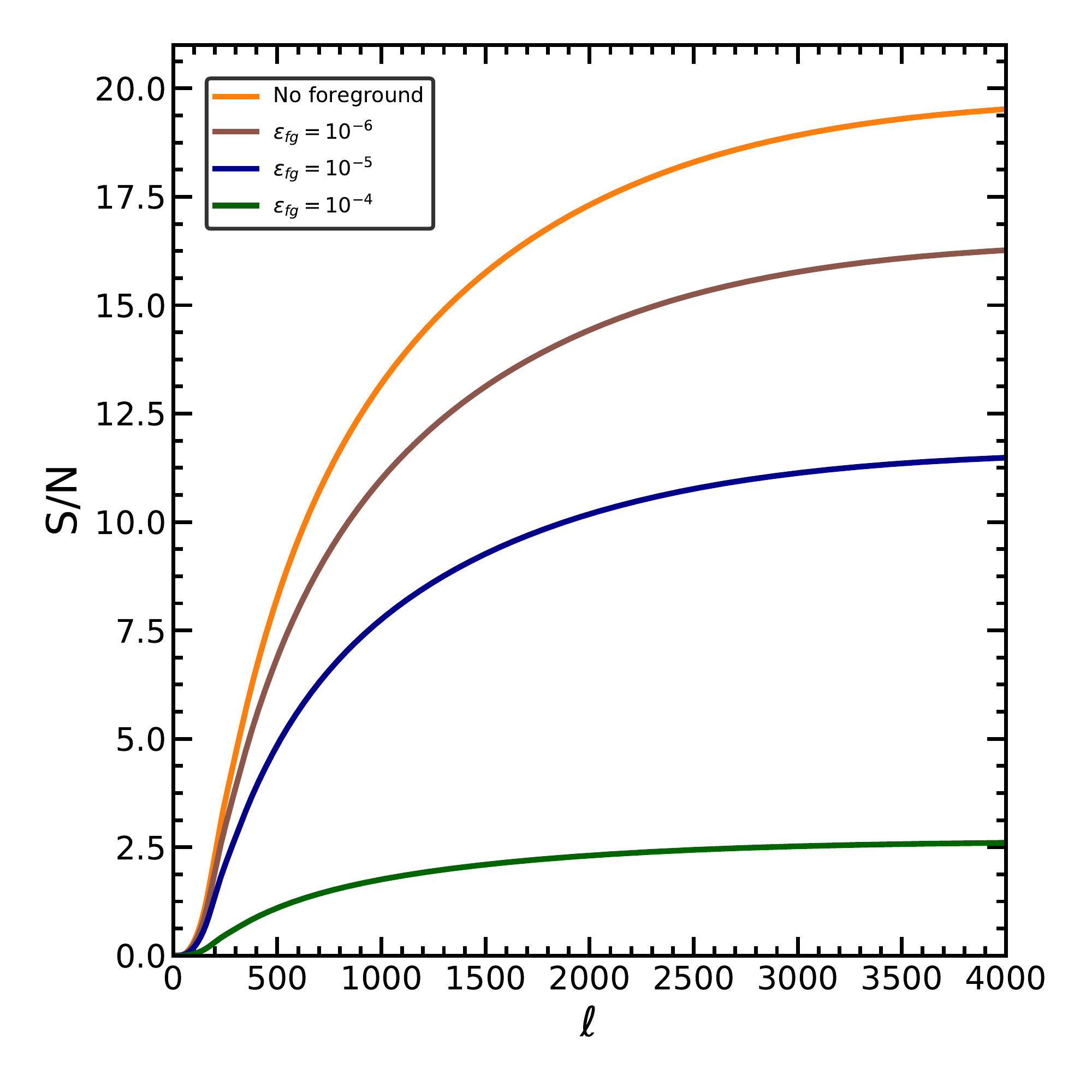}\\
\includegraphics[width=0.6\textwidth]{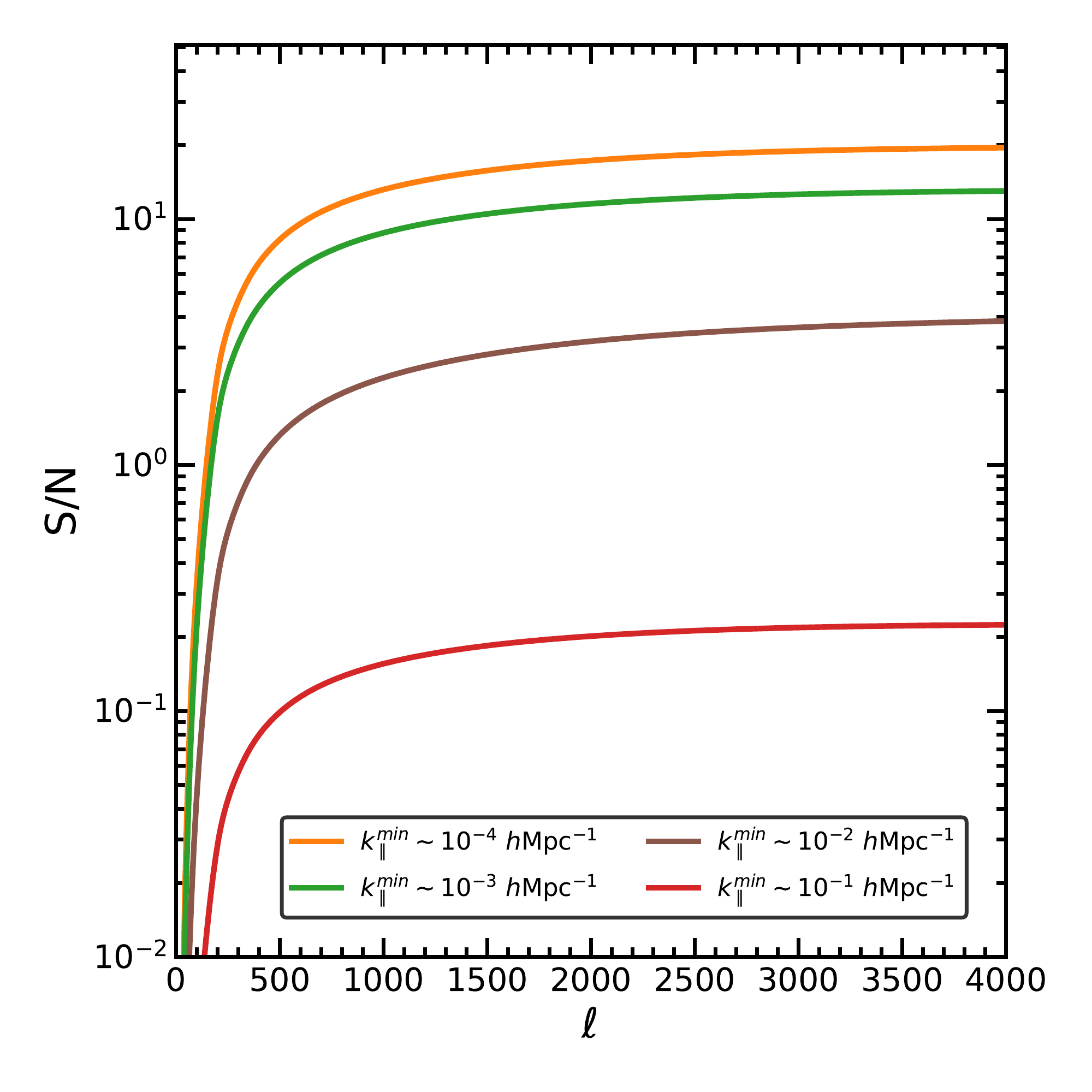}
\caption{Left: we show the effect of residual foreground contamination in 21cm signal on the SNR for the detection of $\tau_{\rm es}-21$cm signal. The orange line represent the SNR for the cross-correlation of PICO$\times$ SKA (as shown in Figure \ref{SNR_ex_comb}). Right: plot for the SNR with the change of minimum value of line of signal wave-number, $k_\parallel$.} 
\label{sn_foreground_fig}
\end{center}
\end{figure}
21cm maps are expected to be dominated by galactic and extra-galactic foregrounds, which may be orders of magnitude larger than the searched cosmological signal. Thus foreground removal will be a crucial task for the extraction of cosmological information from the 21cm observations. In this Section we will briefly discuss how different attempts are being made to characterize the foregrounds contamination and separate them from the cosmological 21cm signal. Foreground mitigation techniques can be classified into two categories: i) foreground removal \cite{2009MNRAS.398..401L,2012MNRAS.423.2518C,2013MNRAS.429..165C, Wang_2013} and ii) foreground avoidance \cite{Datta_2010,Thyagarajan_2015,2015ApJ...809...61A}. 
\begin{table}[t!]
\begin{center}
    \begin{tabular}{| c | c | c | c | c | c | c |}
    \hline
    Foreground & $A^{\rm fg}_{i}$ & $\alpha^i_{\rm fg}$ & $\beta^i_{\rm fg}$ \\
      & [$mK^2$] &  &  \\ \hline
    Synchrotron  & 700 & 2.4 & 2.8 \\ \hline
    point sources  & 57 & 1.1 & 2.07\\ \hline
    free-free & 0.088 & 3.0 & 2.15\\
    \hline
    \end{tabular}
    \caption{Fitted parameters for the 21cm foreground as mentioned in \cite{2005ApJ...625..575S, 2015ApJ...803...21B}.}
   \label{table_foreground}
\end{center}
\end{table}

Foregrounds are expected to be smooth in the frequency domain but to have different spectral properties than the cosmological 21cm signal, hence one can attempt to clean the signal via foreground removal methods. Specifically, in parametric approaches the foreground data are fitted with polynomial coefficients along each line of sight and then are subtracted from the whole dataset in multi-frequency channels which include the cosmological signal. The method works if the foregrounds can be characterized in terms of powerlaws \cite{Wang_2013}. We assume a simple powerlaw model of foregrounds given by \cite{2005ApJ...625..575S,2011MNRAS.413.2103P,2008MNRAS.389.1319J}
\begin{equation}
    C_\ell^{\rm fg}=\epsilon^2_{\rm fg}\sum_i A_i^{\rm fg}\left(\frac{\ell_p}{\ell}\right)^{\alpha^i_{\rm fg}}\left(\frac{\nu_p}{\nu_j}\right)^{\beta^i_{\rm fg}}\,;
\end{equation}

here $A^{\rm fg}$ is the amplitude of the foreground with powerlaw indexes $\alpha_{\rm fg}$ and $\beta_{\rm fg}$. Different values of these parameters are adopted for foregrounds due to synchrotron, free-free and point sources; these are taken from \cite{2005ApJ...625..575S, 2015ApJ...803...21B} and reported for the reader's convenience in Table \ref{table_foreground}. We use $\ell_p=1000$ and $\nu_p=130$ MHz to calculate the angular power spectra of foregrounds. The parameter $\epsilon_{\rm fg}$ represents the foreground removal efficiency, strictly less that unity if a fraction of the foregrounds is removed. 

In the left panel of Figure \ref{sn_foreground_fig}, we show the effect of residual foregrounds on the SNR for detectability of the $\tau_{\rm es}-21$cm signal by varying $\epsilon_{\rm fg}$. In particular, we base on the SNR calculated with PICO$\times$SKA (which was the highest) and investigate how it is affected by residual foreground contamination. In previous Section we considered no residual foreground, which corresponds to the case with $\epsilon_{\rm fg}=0$. When $\epsilon_{\rm fg}$ is increased above $10^{-5}$, foreground contamination becomes comparable to the 21cm signal and the SNR is reduced from 19.5 to 16.2. For $\epsilon_{\rm fg}=10^{-4}$, foreground dominates at all scales, which reduces the SNR significantly to a value around 2. Our basic analysis shows that future 21cm experiment should remove the foreground at the level of $\epsilon_{\rm fg}\approx 7\times10^{-4}$ to detect the cross-correlation signal at a significance of $5\sigma$.

Another important issue to consider is that, being the cosmological signal spread over Fourier space, large-scale line of sight modes ($k_{\parallel}$) will be contaminated due to the smooth nature of foregrounds. The loss of large scale $k_\parallel$ modes could significantly reduce the SNR. Foreground avoidance is the measurement of the power spectra above the minimum value of $k_\parallel$ which is contaminated by foreground. As the spatial model ($k_\perp$) is coupled to the $k_\parallel$ modes, a "wedge" in $k_\perp$-$k_\parallel$ plane is originated in the presence of foregrounds. By avoiding the foreground, one can aim to measure the true power spectrum beyond the wedge, which is known as "EoR window". Rather than working out a detailed analysis of foreground avoidance, which is far beyond the scope of the present paper, we quantitatively study here how the SNR changes if we measure the power spectra above a minimum value of the line of sight  $k^{\rm min}_\parallel$ mode.

To this purpose, we go back to equation \ref{alm21} and decompose $k=\sqrt{|k_\perp|^2+|k_\parallel|^2}$ into $k_\parallel$ and $k_\perp$ modes. In the right panel of Figure \ref{sn_foreground_fig} we show how the loss of line of sight modes due to foreground contamination affects the SNR. Our previous forecasts for the PICO$\times$SKA configuration are recovered when we set $k^{\rm min}_\parallel\approx 10^{-4}\,h\rm Mpc^{-1}$, which essentially mean all of the line of sight modes are included. When $k^{\rm min}_\parallel$ is set to $10^{-3}\,h\rm Mpc^{-1}$ the SNR lowers to values around 12.5. For $k^{\rm min}_\parallel \approx 10^{-1}\,h\rm Mpc^{-1}$ most of the line of sight modes are cut down and the SNR becomes as low as 0.21. Present simulations concerning foreground avoidance suggest that $k^{\rm min}_\parallel\approx 0.03 \,h\rm Mpc^{-1}$ for the fiducial models of foreground by \cite{2014PhRvD..89b3002D}. In that case, the cross--correlation signal could be detected with a significance of about 2.7$\sigma$.

\section{Summary and Outlook}\label{summary}
We have investigated the future detectability of the cross-correlation between fluctuations in the electron scattering optical depth $\tau_{\rm es}$ as probed by CMB experiments, and fluctuations in the 21cm differential brightness temperature $\Delta T_{\rm 21 cm}$ as probed by ground-based radio interferometers. 

Future measurements of the $\tau_{\rm es}-21$cm cross-correlation will probe the evolution of the morphology of the cosmic reionization process, thus shedding light on the properties of the primeval astrophysical sources, and on the distribution of ionized and neutral matter. The $\tau_{\rm es}-21$cm cross-correlation features an inverse bell shape with a minimum at around the multipoles corresponding to the average size of the ionizing bubbles, and a width resulting from the bubble size distribution. The depth of the minimum is maximal at the redshift where the ionizing fraction is around $50\%$ and then becomes shallower at lower and at higher redshifts. The cross power spectrum clearly vanishes in a completely neutral or completely ionized Universe.

We have computed the cumulative SNR expected for the $\tau_{\rm es}-21$cm cross-correlation by combining future CMB experiments probing $\tau_{\rm es}$ (specifically, CMB-S4, PICO and Simons Observatory) with ground-based radio-arrays probing 21cm differential brightness temperature $\Delta T_{\rm 21 cm}$ (specifically, HERA and SKA). We have obtained cumulative SNR larger than 5 for most of the cross configurations, with an optimal SNR around 20 from PICO$\times$SKA. The detectability of the cross spectrum  is weakly dependent on the parameter specifying the reionization morphology (the bubble size distribution), and on the spatially- averaged value of $\tau_{\rm es}$ (at least for $\tau_{\rm es}>0.05$). On the other hand, the SNR is strongly sensitive to the sky fraction $f_{\rm sky}$ commonly covered by CMB and 21cm experiments; for $f_{\rm sky}\sim 1-20\%$, the cumulative SNR increases from values around 5 to about 25. The detailed shape around the minimum can be probed with significances greater than $3\sigma$ only when $f_{\rm sky}$ exceed $10\%$.

Finally, we have discussed how such levels of detectability are affected when (simply modeled) 21cm foregrounds are present. For the most promising PICO$\times$SKA configuration, an efficiency of foreground removal to a level of $7\times 10^{-4}$ is needed to achieve a $5\sigma$ detection of the cross-correlation signal; in addition, safe avoidance of  foreground contamination in the line-of-sight Fourier modes above $0.03 \,h\rm Mpc^{-1}$ would guarantee a detection significance around $3\sigma$.

In the near future measurements of the kSZ signal will potentially be able to probe the morphology of reionization beside putting tighter constrains on the redshift and duration of the reionization process \cite{Smith:2016lnt, 2016ApJ...824..118A,2019arXiv190704473A}. To detect the contribution to the kSZ from patchy reionization, one needs to separate it from the total kSZ signal which is dominated by late-time cosmic structures. This will become feasible with future experiment like CMB S4 via reconstruction of the 2-point and 4-point correlation functions of the kSZ \cite{2019arXiv190704473A,Smith:2018bpn}. Besides, the systematic effects will be different for $\tau_{\rm es}$ fluctuations and kSZ, implying that cross-correlation studies of 21cm fluctuations with kSZ may help to unravel the distribution of ionized bubbles during the reionization process \cite{2010MNRAS.402.2279J, Ma2018}.

Previous works on the $\tau_{\rm es}-21$cm cross-correlation are essentially limited to the reference paper by \cite{Meerburg2013}; with respect to the latter we have adopted here a more realistic reionization model based on the observed high-redshift galaxy luminosity functions (in place of an empirical 'tanh' shape), and gauged on the latest \textit{Planck} 2018 measurements of the integrated optical depth. We stress that our analysis is the first to be focused on the tomographic capability of the $\tau_{\rm es}-21$cm cross-correlation in probing the reionization morphology. We also investigate the impact of 21cm residual foregrounds that can reduce the signal-to-noise ration of the $\tau$-21cm cross-correlation signal. However, the present paper still constitutes a preliminary investigation of such issues and there is plenty of room for further developments, that are certainly needed to strengthen our conclusions. In particular, our future plans include: (i) exploitation of refined algorithms such as excursion set modeling and radiative transfer simulations to describe the distribution of ionized bubbles and its evolution for different reionization histories (Roy et al., in preparation) (ii) more detailed modeling of the foregrounds affecting the 21cm observations; and (iii) application of machine-learning algorithms to quantitatively address the potential of $\tau_{\rm es}-21$cm cross-correlation in reconstructing the parameters describing the astrophysics of primeval ionizing sources and the reionization morphology. 

\acknowledgments{We acknowledge the three anonymous referees for constructive comments that helped to improve the manuscript. We warmly thank Daan Meerburg, Girish Kulkarni and Luigi Danese for useful discussions. This work has been partially supported by PRIN MIUR 2017 prot. 20173ML3WW 002, `Opening the ALMA window on the cosmic evolution of gas, stars and supermassive black holes'. A.L. acknowledges the MIUR grant `Finanziamento annuale individuale attivit\'a base di ricerca' and the EU H2020-MSCA-ITN-2019 Project 860744 `BiD4BEST: Big Data applications for Black hole Evolution STudies'. CB acknowledges support from the INDARK INFN Initiative and the COSMOS Network from the Italian Space Agency (cosmosnet.it)}

\bibliographystyle{apj}
\bibliography{mybib}
\end{document}